\documentclass[onecolumn,a4paper]{article}

\usepackage[english]{babel}
\usepackage{amsthm,amsmath,amssymb,amsfonts,amsopn}
\usepackage{color}
\usepackage{graphicx,subfig,graphics,epstopdf,float}
\graphicspath{{figures/}}
\usepackage{enumerate}
\usepackage{algorithmic}
\usepackage{makecell}
\usepackage{booktabs}
\usepackage{hyperref}
\usepackage{soul}
\setstcolor{red}
\ifpdf
  \DeclareGraphicsExtensions{.eps,.pdf,.png,.jpg}
\else
  \DeclareGraphicsExtensions{.eps}
\fi

\DeclareMathOperator*{\diag}{\mathrm{diag}}

\DeclareMathOperator*{\E}{\mathrm{\bf E}}
\DeclareMathOperator*{\q}{\mathrm{\bf q}}

\newtheorem{thm}{Theorem}
\newtheorem{lem}{Lemma}
\newtheorem{cor}{Corollary}
\newtheorem{assum}{Assumption}
\newtheorem{defn}{Definition}
\newtheorem{rem}{Remark}
\newtheorem{pf}{Proof}

\usepackage{amssymb}
\usepackage{makecell}
\usepackage{color}
\usepackage{subfig,float,graphics}
\graphicspath{{figures/}}
\usepackage{enumerate}
\usepackage{hyperref}

\usepackage{tikz}
\usetikzlibrary{shapes,arrows,automata}
\usetikzlibrary{positioning}
\usetikzlibrary{decorations.pathmorphing} % for snake lines
\usetikzlibrary{matrix,fit} % for block alignment
\usetikzlibrary{calc} % for manipulation of coordinates
\tikzstyle{block} = [draw, rectangle, thick, text centered, minimum height=1cm, minimum width=1.5cm]
\tikzstyle{sum} = [draw, thick, circle, node distance=1cm] % inner sep=0mm, minimum size=2mm,
\tikzstyle{sblock} = [draw, rectangle, thick, text centered, minimum height=0.3cm, minimum width=1.4cm, line width=0.5pt]

\newcommand{\pow}[1]{\|#1\|_{\mathcal{P}}}
\newcommand{\hlt}[1]{#1}

\begin{document}

\title{Optimal output feedback control of \hlt{a class of linear systems with quasi-colored control-dependent multiplicative noise}}
%\tnotemark[1,2]

\author{Junhui Li\thanks{School of Automation Science and Engineering, South China University of Technology,
Guangzhou, China. \href{mailto:wzhsu@scut.edu.cn}{wzhsu@scut.edu.cn},
\href{mailto:l.junhui@mail.scut.edu.cn}{l.junhui@mail.scut.edu.cn},
\href{mailto:aujylu@scut.edu.cn}{aujylu@scut.edu.cn}}
\and Jieying Lu\footnotemark[1]
\and Weizhou Su\footnotemark[1]}

\date{}

\maketitle

\begin{abstract}
This paper addresses the mean-square optimal control problem for \hlt{a class of discrete-time linear systems with a quasi-colored control-dependent multiplicative noise} via output feedback.
The noise under study is novel and shown to have advantage on modeling a class of network phenomena such as random transmission delays.
The optimal output feedback controller is designed using an optimal mean-square state feedback gain and two observer gains, which are determined by the mean-square stabilizing solution to a modified algebraic Riccati equation (MARE), provided that the plant is minimum-phase and left-invertible.
A necessary and sufficient condition for the existence of the stabilizing solution to the MARE is explicitly presented.
It shows that the separation principle holds in a certain sense for the optimal control design of the work.
The result is also applied to the optimal control problems in networked systems with random transmission delays and analog erasure channels, respectively.
\end{abstract}

%\begin{graphicalabstract}
%\includegraphics{figs/grabs.pdf}
%\end{graphicalabstract}
%
%\begin{highlights}
%\item Research highlights item 1
%\item Research highlights item 2
%\item Research highlights item 3
%\end{highlights}

%\begin{keywords}
%optimal control \sep multiplicative noise \sep colored noise \sep networked control system
%\end{keywords}

\section{Introduction}

Recently, stochastic noise is not at all a new phenomenon to the scientific and engineering communities.
The representation of the uncertainties by multiplicative noise has attracted a huge amount of research interests in practical systems.
%If the uncertainty variations concerning a plant or a signal are large, the representation of the uncertainties by multiplicative noise is realistic.
Study in \cite{Elia2005Remote} also shown that stochastic multiplicative noise provides a suitable framework to model communication errors and uncertainties for a number of network phenomena, such as packet drops and network-induced delays.
Hence, control of system with stochastic multiplicative noise has attracted a great deal of research interest.

The mean-square stabilization problem and the optimal control problem are two fundamental issues in studying linear systems with stochastic multiplicative noise.
For independent and identically distributed (i.i.d.) multiplicative noise, the former problem has been studied well by many researchers, see, e.g., \cite{Willems1971Frequency,Lu2002MeanSquare,Qi2017Control} and the references therein.
To mention a few, \cite{Willems1971Frequency} and \cite{Lu2002MeanSquare} provided the well-known mean-square small gain theorems for mean-square stability of linear time-invariant (LTI) single-input and single-output (SISO) systems and multi-input and multi-output (MIMO) systems with i.i.d. multiplicative noise, respectively.
In \cite{Qi2017Control}, fundamental conditions for mean-square stabilizability of generic linear systems with structured i.i.d. multiplicative noise were developed.
Meanwhile, a huge amount of research has been devoted to linear quadratic (LQ) optimal control problem for LTI systems subject to i.i.d. multiplicative noise, see \cite{Wonham1967Optimal,Haussmann1971Optimal,McLane1971Optimal,DEKONING1982Infinite,BEGHI1998Discrete,Huang2008Infinite}.
\hlt{In \cite{Boyd1994Linear}, a linear matrix inequality (LMI) approach is proposed to minimize the upper bound of the LQ performance under unit-energy inputs.}
It is found that the optimal state feedback gain solving the LQ optimal control problem can be obtained by the so-called mean-square stabilizing solution to a modified/generalized algebraic Riccati equation (MARE).
Consequently, most research are focused on providing existence conditions for such mean-square stabilizing solution \cite{Rami2001Solvability,Zhang2008Generalized,Garone2012LQG,Su2016Control,Zheng2018Existence}.
At the early stage, most existing research, e.g., \cite{Rami2001Solvability,Zhang2008Generalized,Garone2012LQG}, provided only sufficient conditions for the existence of the mean-square stabilizing solution, which are often given in terms of some strong conditions such as exact observability of the stochastic system.
In \cite{Su2016Control}, through addressing the mean-square optimal state feedback control problem for a system subject to i.i.d. multiplicative noise, the necessary and sufficient condition for the existence of the mean-square stabilizing solution to the MARE is obtained explicitly, which resolves the longstanding open issue concerning the existence condition for the mean-square stabilizing solution mentioned above.
It is also found that the optimal control problem can be equivalently solved by an LMI-based optimization method.
Moreover, \cite{Su2016Control} also show that the mean-square optimal output feedback control problem also amounts to solving an MARE, provided that the plant is of minimum-phase.
Nowadays, a further result on the existence condition for the mean-square stabilizing solution to the MARE can be found in \cite{Zheng2018Existence}, which is obtained by the theory of cone-invariant operators.
Since i.i.d. multiplicative noise can be applied to model uncertainties of independent lossy and memoryless channels, relevant results on networked control systems (NCSs) are also fruitful, see, e.g., \cite{Braslavsky2007Feedback,SILVA2010Control,Elia2011,Xiao2012Feedback,Zhang2015Linear,Qiu2016}.
Note that the aforementioned results are concern with systems subject to i.i.d. multiplicative noise.
The issue, however, relating to the correlated noise does indeed exist in engineering field and NCSs \cite{Bryson1965Linear,Su2017mean-square,Su2021meansquare}.
For colored stochastic multiplicative noise, much less results on the stability conditions and optimal control are available.
Typically, the colored noise is generated from white noise via a shaping filter, then a differential/difference equation is obtained containing products of the signal in the system with the output of the shaping filter.
Only necessary criteria for the mean-square stability of scalar discrete-time systems were derived in most existing literature \cite{Martin1974Stability,Martin1975Stability,Willems1975Stability}.
Due to the difficulty of the correlation in studying colored noise, results on optimal control for the colored noise system are rarely reported.
Recently, \cite{Li2019Optimal,LI2020Linear} studied a finite-horizon LQ optimal control for a delayed system with colored noise generated by a first-order moving average model with white noise input.
The necessary and sufficient condition for the solvability of optimal control problem was presented by solving the forward and backward stochastic difference equations (FBSDEs) instead of an MARE.
The optimal controller and the associated optimal cost were given in terms of the solution to the FBSDEs, which is complex and difficult to solve even for the first-order colored noise.
Thus, for optimal control problem of linear systems with colored multiplicative noise, there is still a long way to go.

In this work, a \hlt{class of} discrete-time LTI feedback system with a novel stochastic multiplicative noise is studied.
Differing from the white-noise-driven colored noise studied by \cite{Martin1974Stability,Willems1975Stability,Li2019Optimal}, the multiplicative noise under consideration is assumed to be a linear stochastic system with a stationary finite impulse response (FIR), which means that the impulse response of the noise is of finite length and its statistical characteristic is invariant with the time when the impulse applies.
It shows that this kind of multiplicative noise is also an extension of classical i.i.d. noise and has advantage on modeling a class of channel uncertainties induced by random transmission delays and packet dropouts, e.g., analog erasure channel and Rice fading channel \cite{Elia2005Remote}.
\hlt{
Under this framework, an optimal output feedback control problem is addressed.
Notice that optimal output feedback control problem for generic LTI systems with i.i.d. multiplicative noise is non-convex \cite{Xiao2012Feedback,Lu201Mean-square} while the states of left inverse systems can be effectively estimated, the LTI systems considered in this work is assumed to be left-invertible, which significantly simplify the observer gain design and leads to the solvability of the problem.
}
In the same spirit as \cite{Su2016Control}, the optimal controller is designed based on the mean-square stabilizing solution to an MARE, and the necessary and sufficient condition for the existence of such stabilizing solution to the MARE is obtained explicitly.
The associated optimal cost is also given in terms of this solution.
As an application, the stabilization condition of the NCS with analog erasure channel presented in \cite{Elia2005Remote} is recovered.

The remainder of this paper is organized as follows.
Section \ref{Sec:Problem_Formulation} presents the architecture of colored multiplicative noise under study and optimal control problem for a linear feedback system with such noise.
Some preliminary results are also proposed.
Section \ref{Sec:Optimal_control} is devoted to solve the optimal control problem.
Section \ref{Sec:Applications} applies the results to NCSs.
Section \ref{Sec:Simulation} illustrates some numerical examples and Section \ref{Sec:Conclusion} draws conclusions.

The notations used in this paper are mostly standard.
Denote the transpose of a matrix $A$ by $A^T$, the inverse of a nonsingular matrix $A$ by $A^{-1}$, the spectral radius of a matrix $A$ by $\rho(A)$.
Real symmetric matrix $A>0$ (or $\ge 0$) implies that $A$ is strictly positive definite (or positive semi-definite).
The superscript $^*$ represents the complex conjugate transpose.
Denote the expectation of a random variable by $\E\{\cdot\}$, the $\mathcal{H}_2$ norm of a proper stable rational function by $\|\cdot\|_2$ (see \cite{Chen1995Optimal} for details).
$\mathbb{Z}$ and $\mathbb{N}$ are referred to as the set of integers and nonnegative integers, respectively; $\delta(\cdot)$ stands for the Kronecker Delta function.
$\mathcal{F}(\cdot,\cdot)$ is the lower linear fractional transformation \cite{Zhou1995}.

\section{Problem formulation}\label{Sec:Problem_Formulation}

\begin{figure}[hbt]
\centering
\resizebox{0.65\linewidth}{!}{
\begin{tikzpicture}[auto, node distance=3cm, >=stealth', line width=0.6pt]
\node[block, minimum height=1.2cm](P){$P$};
\node[block, below of= P, yshift=0.5cm, minimum width=1.3cm, minimum height=0.8cm](channel){$\Delta$};
\node[block, below right of= P, xshift=0.5cm, yshift=0.8cm, minimum width=1cm, minimum height=0.8cm](K){$K$};
\draw[->](K) |- node[near start]{$u$} (channel);
\draw[->](channel) -- ($(channel)+(-2.6cm,0cm)$) |- node[near end]{$u_d$} ($(P.west)+(0cm,-0.3cm)$);

\draw[->]($(P.west)+(-2.5cm,0.3cm)$) -- node[near start]{$w$} ($(P.west)+(0cm,0.3cm)$);
\draw[->]($(P.east)+(0cm,0.3cm)$) -- node[near end]{$z$} ($(P.east)+(2.5cm,0.3cm)$);
\draw[->]($(P.east)+(0cm,-0.3cm)$) -| node[near start]{$y$} (K);
\end{tikzpicture}
}
\caption{Linear feedback system with multiplicative noise}
\label{Fig:Tracking_System}
\end{figure}
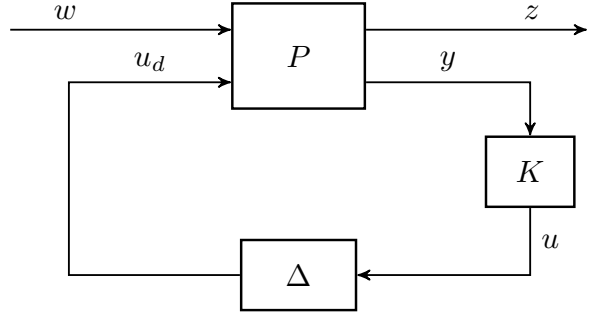

Consider the standard structure of a discrete-time LTI system $P$ with an LTI feedback controller $K$ through a stochastic multiplicative noise $\Delta$, as depicted in Fig. \ref{Fig:Tracking_System}.
The signal $w$ is an external input of the plant $P$; $z$ and $y$ are the controlled output and the measurement of the plant, respectively; $u$ is the controller output, and $u_d$ is the corrupted control signal applied to the plant.
In Fig. \ref{Fig:Tracking_System}, all signals are assumed to be \hlt{scalar-valued except for $y$ and $z$.}
%Note that in this work the aforementioned signals are scalars except $y$.

The state-space model of the plant $P$ is given by
\begin{align}
x(k + 1) &= A x(k) + B_1 w(k) + B_2 u_d(k),~x(0)=0 \nonumber\\
z(k) &= \hlt{C_1 x(k) + D u_d (k)}  \label{Equ:system}\\
y(k) &= C_2 x(k) \nonumber
\end{align}
where $x$ is the state of the plant.
The system matrices in \eqref{Equ:system} have compatible dimensions with the signals.
If the multiplicative noise $\Delta$ has the following properties that for any $k,k_1,k_2\in \mathbb{Z}$,
\begin{equation}\label{Equ:Delta_white}
\begin{aligned}
&\E\{\Delta(k)\}=\mu, \\
&\E\{[\Delta({k_1})-\mu] [\Delta({k_2})-\mu]\} = \delta(k_1-k_2)\sigma^2
\end{aligned}
\end{equation}
where $\Delta(k)$ is referred to as the noise $\Delta$ at instant $k$, then the framework depicted in Fig. \ref{Fig:Tracking_System} is reminiscent of addressing problems about mean-square stabilizability and optimal control of systems with i.i.d. multiplicative noise, see, e.g., \cite{Qi2017Control,Su2016Control}.

Different from \eqref{Equ:Delta_white}, the multiplicative noise $\Delta$ in this work is assumed to be a \emph{causal and linear} stochastic system.
The unit-impulse response of $\Delta$ is given by
\begin{align}\label{Equ:FIR_Delta}
h(k,l) &= \left\{ \begin{array}{ll}
						0, & k<l \\
                        \omega(k,l) ,& k \ge l
                        \end{array} \right.
\end{align}
where $l$ refers to the instant that the impulse applied at, and $\{\omega(k,l): k \ge l\}$ is a sequence of random gains reflecting the stochasticity of $\Delta$.
Consequently, the response of $\Delta$ to its input $u(k)$ is given by the convolution:
\begin{align}
  u_d(k) &= \sum_{l=-\infty}^{\infty} h(k,l) u({l}). \label{Equ:d0}
  %=  \sum_{i \in \mathcal{D}} \omega(k,k-i) u(k-i) .
\end{align}
Throughout this work, we impose the following assumption in finite-length and finite-correlation of the impulse response $h({k,l})$, which causes the noise under consideration to be \emph{quasi-colored} or, simply, colored.

\begin{assum}\label{Assum:omega_k}
The multiplicative noise $\Delta$ is of finite impulse response such that
\begin{align*}
\omega(k,l) = 0,~~\forall ~ k > l + \bar{\tau}
\end{align*}
where $\bar{\tau} \ge 0$ is \hlt{thought} to be the horizon of $\Delta$.
Moreover, denote $\mathcal{D} = \{0,1,\cdots,\bar{\tau}\}$, then
\begin{enumerate}[i)]
\item
for $i \in \mathcal{D}$ and for any $k \in \mathbb{Z}$,
\begin{align}
\hspace{-1cm}\E\{\omega({k+i,k})\} = \mu_i; \label{Equ:Zero-mean}
\end{align}
\item let $\tilde{\omega}({k+i,k})=\omega({k+i,k}) - \mu_i$, then for $i_1,i_2 \in \mathcal{D}$ and $k_1,k_2 \in \mathbb{Z}$,
\begin{align}
\hspace{-1cm}\E\{ \tilde{\omega}({k_1+i_1,k_1}) \tilde{\omega}({k_2+i_2,k_2}) \} =\delta (k_1-k_2) \beta_{i_1,i_2} \label{Equ:Corr-omega}
\end{align}
with $\beta_{i,i} =: \beta_i \ge 0$ and $\beta_{i_1,i_2} = \beta_{i_2,i_1}$ are constants.
\end{enumerate}
\end{assum}

\begin{rem}
Note that the stochastic environment under consideration is characterized only by the first and second order properties of the impulse response sequence of $\Delta$, as shown in Assumption \ref{Assum:omega_k}.
The second item in Assumption \ref{Assum:omega_k} says that $\tilde{\omega}({k_1+i_1,k_1})$ and $\tilde{\omega}({k_2+i_2,k_2})$ are uncorrelated at $k_1 \ne k_2$.
This means that the correlation of the noise only occurs when the noise are corresponding to the same instant when the impulse is applied.
\hlt{Assumption \ref{Assum:omega_k} is quite artificial at first glance.
However, multipath transmission in wireless communication could yield a channel with FIR and similar stochastic properties of the assumption \cite{Goldsmith2005Wireless}.
This implies that we can model delay-induced uncertainties of some communication channels in terms of the memory feature of the noise.
}
\end{rem}

Here are two special cases of the proposed noise applied to networked control systems.
\begin{enumerate}[i)]
  \item When $\bar{\tau}=0$ and $\{\omega({k,k}): k\in \mathbb{Z}\}$ is a white process such that $\mu_0 = \mu$ and $\beta_0 = \sigma^2$, the quasi-colored noise reduces to a memoryless i.i.d. multiplicative noise, which can be used to model uncertainties induced by analog erasure channel \cite{Elia2005Remote}.
  \item When $\bar{\tau}>0$ and, for any given $i\in\mathcal{D}$, the sequence of the random gains $\{\omega({k+i,k}): k\in \mathbb{Z}\}$ is an i.i.d. Gaussian random process with mean $\mu_i$ and variance $\sigma_i^2$, and $\omega(k+i_1,k)$ and $\omega({k+i_2,k})$ are uncorrelated except at $i_1= i_2$, namely $\beta_{i_1,i_2}=0$ for $i_1 \ne i_2$, the noise in this work becomes the stochastic perturbation induced by the Rice fading channel \cite{Elia2005Remote} or the multiple random transmission delays \cite{Li2016Stabilization}.
\end{enumerate}

\subsection{Mean-square stability}

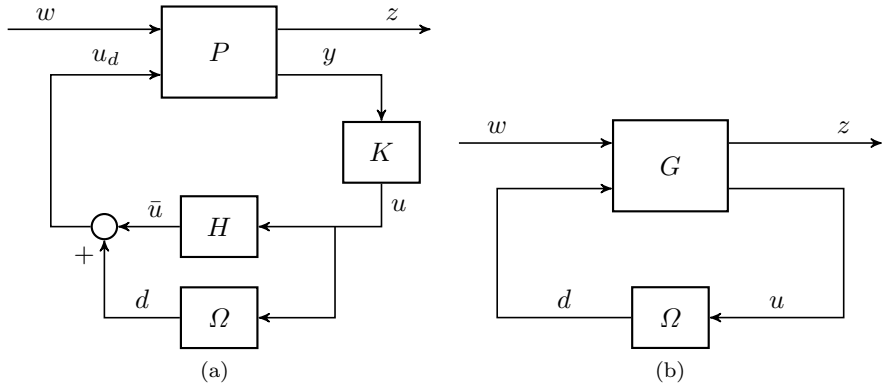
\begin{figure}[hbt]
\centering
\subfloat[]{
\resizebox{0.48\linewidth}{!}{
\begin{tikzpicture}[auto, node distance=3cm, >=stealth', line width=0.6pt]
\node[block, minimum height=1.2cm](P){$P$};
\node[block, below of= P, yshift=0.7cm, minimum width=1cm, minimum height=0.8cm](H){$H$};
\node[sum, left of= H, xshift=-0.5cm](s){};
\node[block, below right of= P, yshift=0.8cm, minimum width=1cm, minimum height=0.8cm](K){$K$};
%\node[block, below left of= P, yshift=0.8cm, minimum width=1cm, minimum height=0.8cm](D){Receiver};
\node[block, below of= H, yshift=1.8cm, minimum width=1cm, minimum height=0.8cm](Omega){$\varOmega$};
\draw[->](K) |- node[near start]{$u$} (H);
\draw[->](H) -- node[pos=0.4, swap]{$\bar{u}$} (s);
\draw[->]($(H.east)+(1cm,0cm)$) |- (Omega);
\draw[->](Omega) -| node[pos=0.9]{$+$} node[near start, swap]{$d$} (s);
\draw[->](s) --++ (-0.7cm,0cm) |- node[near end]{$u_d$} ($(P.west)+(0cm,-0.3cm)$) ;

\draw[->]($(P.west)+(-2cm,0.3cm)$) -- node[near start]{$w$} ($(P.west)+(0cm,0.3cm)$);
\draw[->]($(P.east)+(0cm,0.3cm)$) -- node[near end]{$z$} ($(P.east)+(2cm,0.3cm)$);
\draw[->]($(P.east)+(0cm,-0.3cm)$) -| node[near start]{$y$} (K);
%\draw[->](D) |- node[near end]{$u_d$} ($(P.west)+(0cm,-0.3cm)$);
\end{tikzpicture}
\label{Fig:P_K_H}
}}
\subfloat[]{
\resizebox{0.48\linewidth}{!}{
\begin{tikzpicture}[auto, node distance=2cm, >=stealth', line width=0.6pt]
\node[block, minimum height=1.2cm](G){$G$};
\node[block, below of= G, minimum width=1cm, minimum height=0.8cm](Omega){$\varOmega$};
\draw[->]($(G.west)+(-2cm,0.3cm)$) -- node[near start]{$w$} ($(G.west)+(0cm,0.3cm)$);
\draw[->]($(G.east)+(0cm,0.3cm)$) -- node[near end]{$z$} ($(G.east)+(2cm,0.3cm)$);
\draw[->]($(G.east)+(0cm,-0.3cm)$) -- ++(1.5cm,0cm) |- node[near end, swap]{$u$} (Omega.east);
\draw[->](Omega.west) -| node[near start, swap]{$d$} ($(G.west)+(-1.5cm,-0.3cm)$) -- ($(G.west)+(0cm,-0.3cm)$);
\end{tikzpicture}
}
%\caption{Interconnection of the nominal system and the multiplicative noise induced uncertainty}
\label{Fig:G_with_Omega}
}
\caption{Feedback system over a decomposed noise}
\label{Fig:P_K_H0}
\end{figure}

It is observed that, under Assumption \ref{Assum:omega_k}, the convolution \eqref{Equ:d0} can be rewritten as
\begin{align}
  u_d(k) &= \sum_{i = 0}^{\bar{\tau}}\omega(k,k-i) u(k-i). \label{Equ:d}
\end{align}
Define
\begin{align*}
\bar{u}(k) &= \sum_{i = 0}^{\bar{\tau}} \E\{\omega({k,k-i})\} u(k-i). %\label{Equ:u_hat}
\end{align*}
Since $\E\{\omega({k,k-i})\} = \mu_i$, the signal $\bar{u}(k)$ can be considered as the response of a deterministic linear time-invariant system to the input $u(k)$, i.e.,
\begin{align*}%\label{Equ:H}
H(z) &= \sum_{i=0}^{\bar{\tau}}\mu_i z^{-i},
\end{align*}
which is referred to as the \emph{mean system} of $\Delta$.
Then the residual signal
\begin{align}
d(k) &= u_d(k) - \bar{u}(k) =  \sum_{i = 0}^{\bar{\tau}} \tilde{\omega}({k,k-i}) u(k-i) \label{Equ:d_total}
\end{align}
is with zero-mean and can be considered as the response of a new linear stochastic system, which represents the uncertainty induced by the multiplicative noise and is denoted by $\varOmega$ here.
Accordingly, the corrupted signal $u_d(k)$ is the summation of the outputs of $H$ and $\varOmega$ when considering
$u(k)$ as their inputs, i.e.,
\begin{align*} %\label{Equ:Channel_output}
u_d(k)=\bar{u}(k)+d(k)
\end{align*}
such that the closed-loop system is illustrated by Fig. \ref{Fig:P_K_H}.

Denote by $G$ the closed-loop system from $\{w,d\}$ to $\{z,u\}$ via the controller $K$ when the uncertainty $\varOmega$ is void, i.e.,
\begin{align}\label{Equ:Ge}
G(z) = \begin{bmatrix}
          G_{zw} & G_{zd} \\
          G_{uw} & G_{ud}
        \end{bmatrix}
\end{align}
where $G_{zw}$ maps $w$ to $z$, \emph{et al}.
As it is an uncertainty-free system, $G$ is called the nominal system corresponding to the multiplicative noise system.
Consequently, the closed-loop system can be diagramed as a (lower) linear fractional transformation of the nominal system $G$ with respect to the uncertainty $\varOmega$, as shown in Fig. \ref{Fig:G_with_Omega}.

From \eqref{Equ:d_total}, the (finite) impulse response of $\varOmega$ \hlt{to the unit-impulse signal} is given by
\begin{align*}
\left\{\tilde{\omega}({l,l}),\tilde{\omega}({l+1,l}),\cdots,\tilde{\omega}({l+\bar{\tau},l})\right\}
\end{align*}
where $l$ is the instant when the impulse applies.
\hlt{It is observed from \eqref{Equ:Corr-omega} that $\E\{\tilde{\omega}({l+i,l})\} = 0,\forall i\in \mathcal{D}$ and
\begin{equation}\label{Equ:Property_of_FIR_oemga}
\begin{aligned}
%\E\{\tilde{\omega}({l+i,l})\} &= \mu_{i},\forall i\in \mathcal{D}\\
\E\{\tilde{\omega}({l+i_1,l})\tilde{\omega}({l+i_2,l})\} &= \beta_{i_1,i_2},~~\forall i_1,i_2\in \mathcal{D},
\end{aligned}
\end{equation}
which are independent of the instant $l$.
This implies that the first- and second-order stochastic properties, namely \eqref{Equ:Property_of_FIR_oemga}, of the impulse response of $\varOmega$ is invariant over time and allows us to say that the noise-induced uncertainty $\varOmega$ is stationary in this sense.}
Define the autocorrelation function of the impulse response of $\varOmega$ at $l \in \mathbb{Z}$ as
\begin{align*}
r_l(\lambda)= {\E} \bigg\{\sum_{k=-\infty}^{\infty} \tilde{\omega}({k+\lambda,l}) \tilde{\omega}({k,l})  \bigg\},~~\lambda \in \mathbb{Z}. %\label{Equ:r_Correlation}
\end{align*}
The following lemma is straightforward.\hlt{
\begin{lem}\label{Lma:r} %[\cite{Li2017Stabilizing,Li2018Stability}]
%The noise-induced uncertainty $\varOmega$ is stationary in a certain sense, which means that the second-order stochastic property of the impulse response of $\varOmega$ is time-invariant such that
The autocorrelation function $r_l(\lambda)$ is independent of $l$ and, therefore, can be re-denoted by $r(\lambda)$.
For any $\lambda \in \mathcal{D}$, it holds that
\begin{align*}
r(\lambda) =\sum_{i = 0}^{\bar{\tau}-\lambda} \beta_{i,i+\lambda}
\end{align*}
and $r(-\lambda) = r(\lambda)$; $r(\lambda) = 0$ for $|\lambda| \notin \mathcal{D}$.
\end{lem}}
Subsequently, the energy spectral density of the uncertainty $\varOmega$ can be defined as
\begin{align*} %\label{Equ:Spectral_density}
S(z)=\sum_{\lambda=-\bar{\tau}}^{\bar{\tau}}r(\lambda)z^{-\lambda}.
\end{align*}
Note that $S(z)$ is a spectral density function of a realizable and causal random process, it generally satisfies the Payley-Wiener condition \cite{Papoulis1984Probability}.
That is, there would exist a minimum-phase rational polynomial $\varPhi(z)$ of $\bar{\tau}$th-order such that
\begin{align}\label{Equ:Phi0}
\hlt{S(z)=\varPhi(z)\varPhi(z^{-1})},
\end{align}
which is known as the spectral factorization of $S$, see, e.g., \cite{Zhou1995}, and $\varPhi(z)$ is called the spectral factor of $S(z)$.
Otherwise, the Payley-Wiener condition should be introduced as an assumption of this work.

As the mean system $H(z)$ and spectral factor $\varPhi(z)$ are $\bar{\tau}$th-order polynomials of $z^{-1}$ with real coefficients, they can share the same state space such that
\begin{align}\label{Equ:H_Phi_share}
\begin{bmatrix}
H(z) & \varPhi(z)
\end{bmatrix} = \left[ \begin{array}{c|cc}
                         \hat{A} & {\hat{B}_1} & {\hat{B}_2} \\
                         \hline
                         {\hat{C}} & {\hat{D}_1} & {\hat{D}_2} \\
                       \end{array} \right],
\end{align}
where $\hat{D}_2 = r(0)^{\frac{1}{2}} > 0$.
Note that all eigenvalues of $\hat{A}$ are at the origin.
It is not hard to see that when the colored multiplicative noise reduces to a memoryless multiplicative noise given by the first special case of $\Delta$ in Section \ref{Sec:Problem_Formulation}, it holds
$\begin{bmatrix}
H & \varPhi
\end{bmatrix} = \begin{bmatrix}
\mu & \sigma
\end{bmatrix}$.

Let the set of all proper controllers internally stabilize the nominal system $G(z)$ be $\mathcal{K}_0$.
\begin{defn}\label{Def:Internally_MS}
Suppose the noise $\Delta$ satisfies Assumption \ref{Assum:omega_k}.
The closed-loop system shown in Fig. \ref{Fig:Tracking_System} is mean-square stable if the nominal system $G(z)$ is internally stable, and the variances of the sequences $\{u(k)\}$ and $\{d(k)\}$ are bounded for any i.i.d. input process {$\{w(k)\}$} with bounded variances and being independent of $\Delta$.
\end{defn}

\begin{defn}\label{Def:Mean_square_stabilizable}
The closed-loop system shown in Fig. \ref{Fig:Tracking_System} is mean-square stablilizable if there exists a feedback controller $K \in \mathcal{K}_0$ such that the closed-loop system is mean-square stable.
\end{defn}

The following assumptions on the realization of the plant $P$ are mostly standard in optimal control.
\begin{assum}\label{Assump:A_B_stabilizable}
The plant $P$ satisfies
\begin{enumerate}[i)]
  \item $(A,B_2)$ is stabilizable and $(A,C_1)$ has no unobservable poles on the unit circle; \label{Assump:A_B_stabilizable1}
  \item $(A,C_2)$ is detectable and $(A,[B_1~B_2])$ has no unstabilizable poles on the unit circle. \label{Assump:A_B_stabilizable2}
  \item $H(z) \ne 0$ for any unstable poles of $P(z)$. \label{Assump:A_B_stabilizable3}
\end{enumerate}
\end{assum}

Let $r_1 \ge 1$ and $r_2 \ge 1$ be the relative degrees or input delays of the transfer functions from $w$ and $u_d$ to $y$, respectively.
That is, for $i=1,2$,
\begin{align*}
C_2 A^{j} B_i = 0,~j=0,1,\cdots,r_i-2 \text{ and } C_2 A^{r_i-1} B_i \ne 0.
\end{align*}
Define $G_y = C_2(zI-A)^{-1}\begin{bmatrix}
               B_1 & B_2
              \end{bmatrix}$, the rational matrix from $(w,u_d)$ to $y$, and define $\varPsi = \begin{bmatrix}
                A^{r_1-1}{B}_1 & A^{r_2-1}B_2
              \end{bmatrix}$.
Noticing that the general case is difficult to deal with, we restrict ourselves to minimum-phase and left-invertible case.
\begin{assum}\label{Assum:Relative_degree}
The transfer function $G_y$ satisfies that
\begin{enumerate}[i)]
  \item $G_y$ has no zero outside the unit circle;
  \item the matrix $C_2 \varPsi$ has full column rank.
\end{enumerate}
\end{assum}

\subsection{\texorpdfstring{$\mathcal{H}_2$}{H2} optimal control problem}

Suppose the closed-loop system in Fig. \ref{Fig:G_with_Omega} is mean-square stabilized by the feedback controller $K$.
As the multiplicative noise is stationary, without loss of generality, we assume that the initial time is at $k=0$ and the input sequence $\{w(k): k\in\mathbb{N}\}$ is an {i.i.d.} process with zero-mean and unit-variance.
The mean-square $\mathcal{H}_2$ norm of the system is defined as the square root of the average power of the output signal $z(k)$.
Since this norm is dependent on $K$, denote this average power by $J_{H_2}(K)$, i.e.,
\begin{align}\label{Equ:J_H2}
{J_{{H_2}}}(K) = \|z\|_{\mathcal{P}}^2 := \E\bigg\{ {\mathop {\lim }\limits_{\bar{k} \to \infty } \frac{1}{{\bar{k} + 1}}\sum\limits_{k = 0}^{\bar{k}} \hlt{{z^T(k)}{z(k)} }} \bigg\}
\end{align}
Note that the controlled output $z(k)$ is related to the input noise $w(k)$ and the multiplicative noise $\Delta$, the expectation in \eqref{Equ:J_H2} thus operates jointly over the distributions of $w(k)$ and $\Delta$.

Let $\mathcal{K}$ be the set of all possible controllers \hlt{that} stabilize the closed-loop system in Fig. \ref{Equ:system} in the mean-square sense.
Obviously, $\mathcal{K} \subset \mathcal{K}_0$.
The objective of the $\mathcal{H}_2$ optimal control problem under study is to find an optimal controller $K_{opt}$ to mean-square stabilize the plant \eqref{Equ:system} with multiplicative noise and to minimize the cost $J_{H_2}$, i.e.,
\begin{align*}
{K_{opt}} = \arg \mathop {\inf }\limits_{K \in \mathcal{K}} {J_{{H_2}}}\left( K \right)
\end{align*}
and the corresponding optimal performance is given by
\begin{align*}
%\hspace{-0.5cm}
{J_{opt}} %= \inf_{K \in \mathcal{K}} {J_{{H_2}}}\left( K \right)
= \inf_{K \in \mathcal{K}} \inf_{\sigma_0 > 0} \left\{ {\sigma_0^{-2}}:{J_{{H_2}}}( K ) < {\sigma_0^{-2}}\right\}.
\end{align*}
We have the following characterization of the mean-square $\mathcal{H}_2$ norm, \hlt{which converts the optimal control problem into a stabilizing control problem}.
\begin{lem}\label{Lma:J_H2}
Suppose the feedback system in Fig. \ref{Fig:Tracking_System} satisfies Assumption \ref{Assum:omega_k}.
Then
\begin{enumerate}[i)]
  \item the closed-loop system is mean-square stable if and only if
\begin{align*}
\| G_{ud}(z) \varPhi(z) \|_2^2<1;  % \label{Equ:Ju_PhiG}
\end{align*}
  \item under the mean-square stability condition,
\begin{align*} %\label{Equ:J_H2_norm}
\hspace{-1cm}J_{H_2} = {\left\| {{G_{zw}}} \right\|_2^2 + {{\left\| {{G_{uw}}} \right\|_2^2 \left({1 - \| G_{ud} \varPhi\|_2^2}\right)^{-1} \| G_{zd} \varPhi \|_2^2}}};
\end{align*}
  \item
%\begin{pf}
%See Appendix \ref{Appendix:L7}.
%\end{pf}
%\begin{lem}\label{Lma:Rho}
\hlt{
for any given $\sigma_0 > 0$, that %conditions $J_{H_2}(K) <  {\sigma_0^{-2}}$ and $\|G_{ud}(z)\varPhi(z) \|_2^2<1$ hold
\begin{align*}
J_{H_2}(K) <  {\sigma_0^{-2}}~~\text{and}~~\|G_{ud}(z)\varPhi(z) \|_2^2<1  % \label{Equ:Ju_PhiG}
\end{align*}
if and only if
\begin{align}
\rho(\hat{G})<1
\label{Equ:Total_performance}
\end{align}
where $\hat{G} = \begin{bmatrix}
              \|G_{zw} \sigma_0\|_2^2 &  \| G_{zd}  \varPhi\|_2^2 \\
              \|G_{uw} \sigma_0\|_2^2 &  \| G_{ud}  \varPhi\|_2^2
            \end{bmatrix}$.
            }
\end{enumerate}
\end{lem}

\begin{pf}
See Appendix \ref{Appendix:L7}.
\end{pf}

\section{Mean-square \texorpdfstring{$\mathcal{H}_2$}{H2} optimal control}
\label{Sec:Optimal_control}

%This section is devoted to solve the optimal control problem, where the MARE associated to the optimal control is derived.
%The necessary and sufficient condition of the mean-square stabilizing solution to the MARE is also given explicitly.

\hlt{
According to the last statement of Lemma \ref{Lma:J_H2}, the optimal control problem under study amounts to solving a mean-square stabilizing control problem \eqref{Equ:Total_performance} with a parameter to be optimized.
Since $\hat{G}$ is a positive matrix, the following result is helpful.
}

\begin{lem}[see \cite{Horn1986Matrix}]\label{lem:spectral_radius}
For any $m \times m$ square positive matrix $T = [t_{ij}]$, its spectral radius is
\begin{align*}
  \rho(T) &= \inf_{\varGamma}\max_{1\le j \le m} {\gamma_j^{-2}}\sum_{i=1}^m t_{ij} \gamma_i^2
\end{align*}
where $\varGamma = \diag\{\gamma_1^2,\cdots,\gamma_m^2\} > 0$. % and $t_{ij}$ is the $\{i,j\}$-th entry of the matrix $T$.
\end{lem}

If \eqref{Equ:Total_performance} holds, then by Lemma \ref{lem:spectral_radius} there exists a matrix $\varGamma = \diag\{1,\gamma^2\}>0$ such that
\begin{align}\label{Equ:Norm_in_1}
\|\varGamma \hat{G} \varGamma^{-1} e_i\|_1 < 1,~i=1,2,
\end{align}
where $e_i$ is the $i$th column of the $2 \times 2$ identity matrix.
From the definition of the $\mathcal{H}_2$ norm, the constraints \eqref{Equ:Norm_in_1} can be rewritten as
\begin{align}\label{Equ:Gamma_G1_Gamma_G2}
\sigma_0^2\|\varGamma^{\frac{1}{2}}G_0\|_2^2 < 1,~\|\varGamma^{\frac{1}{2}}{G}_1\|_2^2  {\gamma^{-2}} < 1
\end{align}
where $G_0 = \begin{bmatrix}
        G_{zw}  \\
        G_{uw}
      \end{bmatrix},{G}_1 = \begin{bmatrix}
         G_{zd}  \varPhi\\
        G_{ud} \varPhi
      \end{bmatrix}$.
The set of inequalities in \eqref{Equ:Gamma_G1_Gamma_G2} can be further expressed as
\begin{align*} %\label{Equ:J_H2_bar}
\max_{\substack{\lambda_0>0,\lambda_1>0~\text{s.t.}~\lambda_0^2 + \lambda_1^2 = 1}} J_{\gamma} < 1
\end{align*}
with
\begin{align}\label{Equ:J_gamma}
J_{\gamma} = \lambda_0^2 \sigma_0^2\|\varGamma^{\frac{1}{2}}G_0\|_2^2 + \lambda_1^2\|\varGamma^{\frac{1}{2}}{G}_1\|_2^2 {\gamma^{-2}}.
\end{align}
For any given parameters $\sigma_0,\gamma$ and $\lambda_0,\lambda_1$, direct computation yields that $J_{\gamma} = \|P_\gamma\|_2^2$ where
\begin{align*}
P_\gamma = \varGamma^{\frac{1}{2}} G \diag\{1,\varPhi\} \varPi
\end{align*}
with $\varPi = \begin{bmatrix}
\lambda_0 \sigma_0 & 0\\
0 &  {\lambda_1}{\gamma^{-1}}
\end{bmatrix}$.
\hlt{It can be observed from \eqref{Equ:Gamma_G1_Gamma_G2}-\eqref{Equ:J_gamma} that to solve the equivalent stabilizing problem \eqref{Equ:Total_performance}, the optimal control for plant $P_\gamma$ via output feedback controller $K$ should be considered first.
By the following result, the later problem is solved standardly.
}

\begin{lem}\label{Lma:P_gamma_order_reduced_realization}
It holds that
\begin{align*}
\|P_\gamma\|_2^2 = \|\mathcal{F}(\bar{P}_\gamma,K)\|_2^2
\end{align*}
where the auxiliary plant $\bar{P}_\gamma$ admits the following realization:
\begin{align}
\bar{x}(k+1) &= \bar{A} \bar{x}(k) \!+\! \begin{bmatrix}
                                            \bar{B}_{1} & \bar{B}_{2}
                                          \end{bmatrix} \varPi \begin{bmatrix}
                                                          v(k) \\
                                                          \varphi(k)
                                                        \end{bmatrix} \!+\! \tilde{B}_{2} u(k)\nonumber\\
\bar{z}(k) &= \bar{C}_{1} \bar{x}(k) + \hlt{\bar{D}_{11}\varPi  \begin{bmatrix}
                                                          v(k) \\
                                                          \varphi(k)
                                                        \end{bmatrix} } + \bar{D}_{\gamma} u(k) \label{Equ:Auxiliary_Plant_1} \\
\bar{y}(k) &= \bar{C}_{2} \bar{x}(k) \nonumber
\end{align}
with $\bar{A}\!= \!\left[\begin{smallmatrix}
     A & B_2 \hat{C} \\
      0 & \hat{A}
    \end{smallmatrix}\right],\bar{B}_{1} \!=\! \left[\begin{smallmatrix}
         B_1 \\
         0\\
       \end{smallmatrix}\right],\bar{B}_{2}\! =\! \left[\begin{smallmatrix}
         B_2 \hat{D}_2 \\
          \hat{B}_2
       \end{smallmatrix}\right],
\tilde{B}_{2}\! =\! \left[\begin{smallmatrix}
         B_2 \hat{D}_1 \\
         \hat{B}_1
       \end{smallmatrix}\right]$, $
\bar{C}_{1}\! = \!\left[\begin{smallmatrix}
       C_1 & \hlt{D \hat{C}}\\
       0 & 0
       \end{smallmatrix}\right],
\bar{C}_{2} \!= \!\left[\begin{smallmatrix}
       C_2 & 0
       \end{smallmatrix}\right],
\hlt{\bar{D}_{11} \! = \!\left[\begin{smallmatrix}
       0 & D \hat{D}_2\\
       0 & 0
       \end{smallmatrix}\right] ,
\bar{D}_{\gamma} \!= \!\left[\begin{smallmatrix} D \hat{D}_1 \\ \gamma \end{smallmatrix}\right]}$.
\end{lem}

\begin{pf}
See Appendix \ref{Appendix:P_gamma}.
\end{pf}
\begin{rem}
Note that the auxiliary plant $\bar{P}_\gamma$ is composed of the plant $P$, the mean system $H$, and the uncertainty spectral factor $\varPhi$, while the order of its realization in \eqref{Equ:Auxiliary_Plant_1} is only the sum of the order of $P$ and the memory length of the colored multiplicative noise.
This is sensible since both $H$ and $\varPhi$ are derived from one multiplicative noise such that they can physically share the same state-space, as shown in \eqref{Equ:H_Phi_share}.
\end{rem}

%\begin{lem}\label{Lem:AB_AbarBbar}
%Assumption \ref{Assump:A_B_stabilizable}.1 and \ref{Assump:A_B_stabilizable}.2 are equivalent to the following statements, respectively,
%\begin{enumerate}[i)]
%  \item $(\bar{A},\tilde{B}_{2})$ is stabilizable and $(\bar{A},\bar{C}_{1})$ has no unobservable poles on the unit circle;
%  \item $(\bar{A},\bar{C}_{2})$ is detectable and $(\bar{A},[\bar{B}_{1}~\bar{B}_{2}])$ has no unstabilizable poles on the unit circle.
%\end{enumerate}
%\end{lem}
%\begin{pf}
%See Appendix \ref{Appendix:AB_AbarBbar}.
%\end{pf}

%According to Lemma \ref{Lem:AB_AbarBbar}, the optimal design for the plant $P_\gamma$ is solvable.
For any given $\sigma_0,\lambda_0,\lambda_1$, and $\gamma$, the optimal output feedback controller for plant \eqref{Equ:Auxiliary_Plant_1} is solved from the stabilizing solutions $X_\gamma$ and $Y_\gamma$ to the following two discrete-time algebraic Riccati equations (DAREs), respectively,
\begin{align}
X_{\gamma} &= \bar{A}^T X_{\gamma} \bar{A} + \bar{C}_{1}^T \bar{C}_{1} - {(\bar{A}^T X_{\gamma}{\tilde{B}_{2}} + \bar{C}_{1}^T \hlt{\bar{D}_{12}} )} \nonumber\\
&\hspace{0cm}\times (M_{\gamma} +  {\tilde{B}_{2}^T}X_{\gamma}{\tilde{B}_{2}})^{-1} ({\tilde{B}_{2}^T}X_{\gamma} \bar{A} + \hlt{\bar{D}_{12}^T} \bar{C}_{1}) \label{Equ:Riccati_Equ_X}
\end{align}
and
\begin{align}
Y_{\gamma} &= \bar{A} Y_{\gamma} \bar{A}^T + {\bar{B}_{1}}\sigma_0^2 \lambda_0^2 {{\bar{B}_{1}^T}} +  \bar{B}_{2} {\lambda_1^2}{\gamma^{-2}}\bar{B}_{2}^T  \nonumber \\
&\hspace{2cm}-  \bar{A} Y_{\gamma}{{\bar{C}_{2}^T}}{( {{{\bar{C}_{2}}}Y_{\gamma}{{\bar{C}_{2}^T}}} )^\dag }{\bar{C}_{2}}Y_{\gamma} \bar{A}^T  \label{Equ:Riccati_Equ_Y}
\end{align}
where $M_\gamma \!=\! \bar{D}_{12}^T \bar{D}_{12} + \gamma^2$ with $\bar{D}_{12} \!=\!  \left[ \begin{smallmatrix} \hlt{D \hat{D}_1} \\ 0 \end{smallmatrix}\right]$, and $\bar{C}_{1}^T \bar{D}_{12} \!=\! \bar{C}_{1}^T \bar{D}_{\gamma}$.

\begin{lem}\label{Lma:Optimal_for_auxiliary_plant}
Under Assumptions \ref{Assump:A_B_stabilizable}, the DAREs \eqref{Equ:Riccati_Equ_X} and \eqref{Equ:Riccati_Equ_Y} have unique stabilizing solutions $X_{\gamma}$ and $Y_{\gamma}$, respectively.
In addition, $X_{\gamma}$ and $Y_{\gamma}$ are positive semi-definite.
The $\mathcal{H}_2$ optimal controller for the auxiliary plant \eqref{Equ:Auxiliary_Plant_1} is
\begin{align*} %\label{Equ:Optimal_Controller}
&K_{\gamma}(z) = \\
&\left[ {\begin{array}{c|c}
{\bar{A} + {\tilde{B}_{2}}{F_{\gamma}}+ L_{\gamma}{{\bar{C}_{2}}} - {\tilde{B}_{2}}{L_{0\gamma}}{\bar{C}_{2}}}&{{\tilde{B}_{2}}{L_{0\gamma}} - L_{\gamma}}\\
\hline
{{F_{\gamma}}- {L_{0\gamma}}{\bar{C}_{2}}}&{{L_{0\gamma}}}
\end{array}}\right]
\end{align*}
where
\begin{align}\label{Equ:Optimal_F}
F_{\gamma} = - (M_{\gamma} +  {\tilde{B}_{2}^T}X_{\gamma}{\tilde{B}_{2}})^{-1}({{{\tilde{B}_{2}^T}X_{\gamma} \bar{A}} + \bar{D}^T \bar{C}_{1} })
\end{align}
and
\begin{align*}
L_{\gamma} \!=\!  -  {\bar{A} Y_{\gamma}{{\bar{C}_{2}^T}}} {( {{\bar{C}_{2}}Y_{\gamma} {\bar{C}_{2}^T}} )^{\dagger}},~ %\label{Equ:Optimal_L}\\
{L_{0\gamma}} \!=\!  {{F_{\gamma}} Y_{\gamma}{{\bar{C}_{2}^T}}} {( {{\bar{C}_{2}}Y_{\gamma}{\bar{C}_{2}^T}} )^{\dagger}} %\label{Equ:Optimal_L0}
\end{align*}
The associated minimum performance cost is
\begin{align*}
&\hspace{-0.7cm}\min_{K \in \mathcal{K}}J_{\gamma} = \sigma_0^2 \lambda_0^2{\bar{B}_{1}^T}{X_{\gamma}}{\bar{B}_{1}} + {\lambda_1^2}{\gamma^{-2}} (\bar{B}_{2}X_{\gamma} \bar{B}_{2}^T \hlt{+ \hat{D}_2^T D^T D \hat{D}_2})   \nonumber\\
&\hspace{0.5cm}+R_\gamma(L_{0\gamma}\bar{C}_{2} - F_{\gamma})Y_{\gamma}(L_{0\gamma}\bar{C}_{2} - F_{\gamma})^T R_\gamma^T
%\label{Equ:Optimal_J_gamma}
\end{align*}
where $R_\gamma = (M_\gamma + \tilde{B}_{2}^T X_{\gamma} \tilde{B}_{2})^{\frac{1}{2}}$.
\end{lem}

\begin{pf}
%The existence and uniqueness of the stabilizing positive semi-definite solutions to the DAREs \eqref{Equ:Riccati_Equ_X} and \eqref{Equ:Riccati_Equ_Y} are from Lemma \ref{Lem:AB_AbarBbar}, while the rest is standard for discrete-time LTI systems, see \cite{Chen1995Optimal}.
See Appendix \ref{Appendix:AB_AbarBbar}.
\end{pf}

Since the DARE \eqref{Equ:Riccati_Equ_Y} is related to the parameters $\lambda_0$ and $\lambda_1$, the optimal parameters $L_\gamma$ and $L_{0\gamma}$ are strongly coupled with $\lambda_0$ and $\lambda_1$.
This makes the optimal design problem difficult to solve.
However, Assumption \ref{Assum:Relative_degree} yields that the stabilizing solution to the DARE \eqref{Equ:Riccati_Equ_Y} can be simplified and given explicitly such that the optimal parameters $L_\gamma$ and $L_{0\gamma}$ are decoupled from $\lambda_0$ and $\lambda_1$.

%\begin{lem}[see \cite{Silverman1976Discrete}]\label{Lem:Y_gamma}
%Suppose the plant \eqref{Equ:system} satisfies Assumptions \ref{Assump:A_B_stabilizable} and \ref{Assum:Relative_degree}, then for given $\lambda_0,\lambda_1$, and $\gamma$, the stabilizing solution to the DARE \eqref{Equ:Riccati_Equ_Y} is given by
%\begin{align*} %\label{Equ:Y}
%Y_{\gamma} \!=\! \sum_{i=0}^{r_1 - 1}\bar{A}^i \bar{B}_{1}\sigma_0^2 \lambda_0^2 \bar{B}_{1}^T \bar{A}^{T^i} \!+\! \sum_{i=0}^{r_2 - 1}\bar{A}^i \bar{B}_{2}  {\lambda_1^2}{\gamma^{-2}} \bar{B}_{2}^T \bar{A}^{T^i}.
%\end{align*}
%\end{lem}

\begin{lem}\label{Lma:L_L0_J}
Suppose that  Assumptions \ref{Assump:A_B_stabilizable} and \ref{Assum:Relative_degree} hold, then, for the auxiliary plant \eqref{Equ:Auxiliary_Plant_1}, the $\mathcal{H}_2$ optimal output feedback parameters are given by \eqref{Equ:Optimal_F} and
\begin{align}
L &=  -  {\bar{A} \bar{\varPsi}} {( \bar{C}_{2}\bar{\varPsi} )^{\dagger}},~~{L_{0\gamma}} =  {{F_{\gamma}}\bar{\varPsi}} {( \bar{C}_{2}\bar{\varPsi} )^{\dagger}} \label{Equ:Optimal_L1}
\end{align}
where $\bar{\varPsi} = \begin{bmatrix}
               \bar{A}^{r_1-1}\bar{B}_{1} & \bar{A}^{r_2-1}\bar{B}_{2}
              \end{bmatrix}$.
Moreover, the minimum performance cost of the resulting closed-loop system is given by
\begin{align}
&\min_{K}J_{\gamma} = \lambda_0^2 \sigma_0^2  \phi_0(X_{\gamma}) + {\lambda_1^2}{\gamma^{-2}}\phi_1(X_{\gamma}) \label{Equ:Optimal_J_1}
\end{align}
where
\begin{align*}
\hspace{-0.8cm}\phi_0(X_{\gamma}) = {\bar{B}_{1}^T}\bar{A}^{T^{r_1-1}}{X_{\gamma}} \bar{A}^{{r_1-1}} {\bar{B}_{1}}+\sum_{j=0}^{r_1- 2}{\bar{B}_{1}^T}\bar{A}^{T^{j}}\bar{C}_{1}^T \bar{C}_{1} \bar{A}^{j} {\bar{B}_{1}}
\end{align*}
and
\begin{align}
&\hspace{-0.8cm}\phi_1(X_{\gamma}) = {\bar{B}_{2}^T}\bar{A}^{T^{r_2-1}}{X_{\gamma}} \bar{A}^{{r_2-1}} {\bar{B}_{2}} +\sum_{j=0}^{r_2 - 2}{\bar{B}_{2}^T}\bar{A}^{T^{j}}\bar{C}_{1}^T \bar{C}_{1} \bar{A}^{j} {\bar{B}_{2}}\nonumber\\
&\hspace{1cm}\hlt{ + \hat{D}_2^T D^T D \hat{D}_2} \label{Equ:Phi_1X}
\end{align}
\end{lem}

\begin{pf}
See Appendix \ref{Appendix:L_L0}.
\end{pf}

By comparing \eqref{Equ:Optimal_J_1} with \eqref{Equ:J_gamma}, it is observed that under the optimal controller parameters $F_\gamma,L,L_{0\gamma}$ and for the given $\gamma$,
\begin{align*}
\|\varGamma^{\frac{1}{2}}{G}_0\|_2^2 = \phi_0(X_{\gamma}),~\|\varGamma^{\frac{1}{2}}{G}_1\|_2^2 = \phi_1(X_{\gamma}).
\end{align*}
Thus, the design constraints in \eqref{Equ:Gamma_G1_Gamma_G2} for the mean-square $\mathcal{H}_2$ optimal control problem are written as
\begin{align}\label{Equ:Constraint_psi}
\sigma_0^2\phi_0(X_{\gamma})<1~~\text{and}~~ {\gamma^{-2}}\phi_1(X_{\gamma}) < 1.
\end{align}
Note that the optimal parameters are not related to $\lambda_0$ and $\lambda_1$ but only dependent on the state feedback gain $F_\gamma$ and the auxiliary plant.
This implies that now the key issue in the mean-square optimal output feedback design is to find the maximum $\sigma_0$, the optimal state feedback $F_\gamma$, and the associated $\gamma$ subject to the design constraint \eqref{Equ:Constraint_psi}.

To this end, we study the mean-square $\mathcal{H}_2$ optimal state feedback problem for a new auxiliary plant with a \emph{white} multiplicative noise.
That is,
\begin{align}
\tilde{x}(k+1) &= \bar{A} \tilde{x}(k) \!+\! \bar{A}^{r_1-1}\bar{B}_{1} v(k) \nonumber\\
&\hspace{2cm}+ \bar{A}^{r_2-1} \bar{B}_{2} \varphi(k) + \tilde{B}_{2} u(k) \nonumber\\
\tilde{z}(k) &= \bar{C}_1 \tilde{x}(k) \!+\!   D_{1}  v(k) \!+\!  D_2 \varphi(k) \!+\! \bar{D}  u(k) \label{Equ:Plant_IID_II}
\end{align}
with $\varphi(k) = \bar{\Delta}(k) u(k)$, where $\{\bar{\Delta}(k): k\in \mathbb{Z}\}$ is an {i.i.d. white noise process} with zero-mean and unit-variance, and $D_1,D_2$ satisfy
\begin{align*}
D_1^T D_1 &= \sum_{j=0}^{r_1- 2}{\bar{B}_{1}^T}\bar{A}^{T^{j}}\bar{C}_{1}^T \bar{C}_{1} \bar{A}^{j} {\bar{B}_{1}}\\
D_2^T D_2&= \sum_{j=0}^{r_2- 2}{\bar{B}_{2}^T}\bar{A}^{T^{j}}\bar{C}_{1}^T \bar{C}_{1} \bar{A}^{j} {\bar{B}_{2}} \hlt{ + \hat{D}_2^T D^T D \hat{D}_2}.
\end{align*}
By \cite[Lemma 5]{Su2016Control}, the problem of mean-square $\mathcal{H}_2$ optimal control for the plant \eqref{Equ:Plant_IID_II} via \emph{state feedback} is to find $F_{opt} = \arg\inf_{F}\inf_{\sigma_0}\sigma_0^{-2}$ subject to
\begin{align}
{\sigma_0^{2}}\|\varGamma^{\frac{1}{2}}\bar{G}_0 \|_2^2 < 1 ,~\gamma^{-2}\|\varGamma^{\frac{1}{2}}\bar{G}_1 \|_2^2 < 1,
\label{Equ:Constratin_G_bar}
\end{align}
where, for $i=1,2$,
\begin{align*}
\bar{G}_i &= \begin{bmatrix}
                                                  C_{F}\\
                                                  F
                                                \end{bmatrix}(zI-\bar{A} - \tilde{B}_{2}F)^{-1}\bar{A}^{r_i - 1}\bar{B}_i + \begin{bmatrix}
                                                                                                      D_i \\
                                                                                                      0
                                                                                                    \end{bmatrix}
\end{align*}
with $\varGamma = \diag\{1,\gamma^2\}$ and $C_F =  \bar{C}_{1} + \bar{D}F$, as the multiplicative noise in \eqref{Equ:Plant_IID_II} is of unit-variance.
On the other hand, for the given $\gamma$, it is well-known (see, e.g., \cite{Chen1995Optimal}) that $F_\gamma$ given in \eqref{Equ:Optimal_F} simultaneously minimizes $\|\varGamma^{\frac{1}{2}}\bar{G}_0 \|_2^2$ and $\|\varGamma^{\frac{1}{2}}\bar{G}_1 \|_2^2$ such that
\begin{align*}
\|\varGamma^{\frac{1}{2}}\bar{G}_0 \|_2^2  = \phi_0(X_\gamma) ,~\|\varGamma^{\frac{1}{2}}\bar{G}_1 \|_2^2 = \phi_1(X_\gamma),
\end{align*}
where $X_\gamma$ is the stabilizing solution to the DARE \eqref{Equ:Riccati_Equ_X}.
As a result, the output feedback optimization problem for $\bar{P}_{\gamma}$ is converted into a state feedback one for \eqref{Equ:Plant_IID_II}, while the latter can be solved by the following result.

%\begin{lem}[see \cite{Su2016Control}]\label{Lma:DARE_X}
%The following MARE
%\begin{align}
%&X = \bar{A}^T X \bar{A} + \bar{C}_{1}^T \bar{C}_{1} - (\bar{A}^T X {\tilde{B}_{2}} + \bar{C}_{1}^T \hlt{\bar{D}_{12}}) \nonumber\\
%&\hspace{0.3cm}\times [{{M(X) + {\tilde{B}_{2}^T}X{\tilde{B}_{2}}}}]^{-1} ({\tilde{B}_{2}^T}X \bar{A} + \hlt{\bar{D}_{12}^T} \bar{C}_{1}) , \label{Equ:DARE_X_II}
%\end{align}
%where $M(X) = \phi_1(X) + \hlt{\bar{D}_{12}^T \bar{D}_{12} } $ with $\phi_1(X)$ in \eqref{Equ:Phi_1X}, has a mean-square stabilizing solution if and only if the plant \eqref{Equ:Plant_IID_II} is mean-square stabilizable and $(A,C_1)$ has no unobservable poles on the unit circle.
%In addition, this solution is the unique mean-square stabilizing solution and the largest positive semi-definite solution\footnote{\hlt{$X$ being the largest semi-definite solution means that $X \ge \tilde{X}$ for any other solution $\tilde{X}$.}} to the MARE \eqref{Equ:DARE_X_II}.
%\end{lem}

\begin{lem}\label{Lma:Optimal_F_J}
Suppose $(A,C_1)$ has no unobservable poles on the unit circle.
Then the plant \eqref{Equ:Plant_IID_II} is mean-square stabilizable if and only if the largest solution \footnote{\hlt{$X$ being the largest semi-definite solution means that $X \ge \tilde{X}$ for any other solution $\tilde{X}$.}} $X$ to the following MARE is positive semidefinite:
\begin{align}
&X = \bar{A}^T X \bar{A} + \bar{C}_{1}^T \bar{C}_{1} - (\bar{A}^T X {\tilde{B}_{2}} + \bar{C}_{1}^T \hlt{\bar{D}_{12}}) \nonumber\\
&\hspace{0.3cm}\times [{{M(X) + {\tilde{B}_{2}^T}X{\tilde{B}_{2}}}}]^{-1} ({\tilde{B}_{2}^T}X \bar{A} + \hlt{\bar{D}_{12}^T} \bar{C}_{1}) \label{Equ:DARE_X_II}
\end{align}
where $M(X) = \phi_1(X) + \hlt{\bar{D}_{12}^T \bar{D}_{12} } $ with $\phi_1(X)$ in \eqref{Equ:Phi_1X}.
When the above holds, the mean-square optimal $\mathcal{H}_2$ state feedback gain is given by
\begin{align}
&F = - [M(X) +  {\tilde{B}_{2}^T}X{\tilde{B}_{2}}]^{-1}({{{\tilde{B}_{2}^T}X \bar{A}} + \bar{D}^T \bar{C}_{1} }), \label{Equ:Optimal_F_II}
\end{align}
and the associated minimum performance cost for the resulting closed-loop system is given by
\begin{align}
\min_{F}\tilde{J}_{H_2} &=  {{B}_{1}^T}A^{T^{r_1-1}}{X_{11}} A^{{r_1-1}} {{B}_{1}} \nonumber \\
&\hspace{1.5cm}  +\sum_{j=0}^{r_1- 2}{{B}_{1}^T}A^{T^{j}}C_{1}^T C_{1} A^{j} {{B}_{1}} \label{Equ:Optima_Jh2_1}
\end{align}
where $X_{11} \ge 0$ is the upper-left corner of $X$ with compatible dimension of $A$.
\end{lem}

\begin{pf}
See Appendix \ref{Appendix:Jh2_F}.
\end{pf}

\hlt{
Note that the equivalent stabilizing problem is solved by Lemma \ref{Lma:Optimal_for_auxiliary_plant}-\ref{Lma:Optimal_F_J}, and the parameter $\sigma_0^{-2}$ is simultaneously minimized by \eqref{Equ:Optimal_F_II}.
It is ready to state the main result of this work. %which solves the mean-square optimal output feedback control problem to the plant \eqref{Equ:system} with colored multiplicative noise.
}

\begin{thm}\label{Thm:K_opt_L_L0}
Suppose the plant \eqref{Equ:system} with a colored multiplicative noise $\Delta$ satisfies Assumptions \ref{Assum:omega_k}, \ref{Assump:A_B_stabilizable}, and \ref{Assum:Relative_degree}.
\begin{enumerate}[i)]
  \item The MARE \eqref{Equ:DARE_X_II} has a mean-square stabilizing solution if and only if the plant \eqref{Equ:system} is mean-square stabilizable and $(A,C_1)$ has no unobservable poles on the unit circle.
  \item If the MARE \eqref{Equ:DARE_X_II} has a mean-square stabilizing solution $X \ge 0$, then the minimum performance cost $J_{H_2}$ of the plant \eqref{Equ:system} is given by \eqref{Equ:Optima_Jh2_1}, with the associated optimal output feedback controller given by
\begin{align}
\hspace{-1.4cm}K_{opt}(z) = \left[ {\begin{array}{c|c}
{\bar{A} \!+\! {\tilde{B}_{2}}{F}\!+\! L{{\bar{C}_{2}}} \!-\! {\tilde{B}_{2}}{L_0}{\bar{C}_{2}}}&{{\tilde{B}_{2}}{L_0} \!-\! L}\\
\hline
{{F}- {L_0}{\bar{C}_{2}}}&{{L_0}}
\end{array}}\right] \label{Equ:Kopt}
\end{align}
where the optimal parameters are given by \eqref{Equ:Optimal_F_II} and
\begin{align}
L &=  -  {\bar{A} \bar{\varPsi}} {( \bar{C}_{2}\bar{\varPsi} )^{\dagger}},~~{L_0} =  {{F}\bar{\varPsi}} {( \bar{C}_{2}\bar{\varPsi} )^{\dagger}}. \label{Equ:Optimal_Ls}
\end{align}
\end{enumerate}
\end{thm}

\begin{pf}
If the mean-square stabilizability of the plant \eqref{Equ:system} and that of the associated auxiliary plant \eqref{Equ:Plant_IID_II} are equivalent, then, by Lemma \ref{Lma:Optimal_F_J}, the first statement is established.
To see the equivalence, suppose the former plant is mean-square stabilizable, then there exists a mean-square stabilizing controller, say $K$, and a sufficiently small $\sigma>0$ such that \eqref{Equ:Total_performance} holds.
Consequently, \eqref{Equ:Gamma_G1_Gamma_G2} holds for some $\gamma>0$.
Since the controller $K_\gamma$ given by Lemma \ref{Lma:Optimal_for_auxiliary_plant} and Lemma \ref{Lma:L_L0_J} is superior to $K$, in the sense of minimizing $J_\gamma$ with {any} given $\lambda_0,\lambda_1$ satisfying $\lambda_0^2+\lambda_1^2=1$, \eqref{Equ:Constratin_G_bar} holds for $F_\gamma$ given by \eqref{Equ:Optimal_F}.
Here it is sufficient to see that the plant \eqref{Equ:Plant_IID_II} is mean-square stabilizable.
The converse is quite similar so that is omitted.

For the second statement, as mentioned above, the optimal output feedback design is equivalent to find $F_\gamma$ in minimizing $\sigma_0^{-1}$ subject to the constraints
\begin{align*}
\sigma_0^2\phi_0(X_{\gamma})<1~~\text{and}~~ {\gamma^{-2}}\phi_1(X_{\gamma}) < 1,
\end{align*}
where $X_\gamma$ is the stabilizing solution to the DARE \eqref{Equ:Riccati_Equ_X}.
By applying Lemma \ref{Lma:Optimal_F_J}, we obtain the optimal parameter $F$ given by \eqref{Equ:Optimal_F_II} for the output feedback and the minimum performance cost $\phi_0(X)$, where $X$ is the mean-square stabilizing solution to the MARE in Lemma \ref{Lma:Optimal_F_J}.
Then the proof is established.
\end{pf}

\begin{rem}
\hlt{
%The mean-square stabilizability condition of the plant \eqref{Equ:Plant_IID_II} can be seen in \cite{Qi2017Control}.
%Note that the order of the optimal controller \eqref{Equ:Kopt} is the sum of the plant's order and the memory length of the multiplicative noise.
%The optimal state feedback gain is obtained from the mean-square stabilizing solution to the MARE \eqref{Equ:DARE_X_II}, and the remaining parameters $L$ and $L_0$, which are related to the observer-like part of the controller, are obtained from the plant, the noise model, and the optimal state feedback gain.
%Thus,
Theorem \ref{Thm:K_opt_L_L0} is reminiscent of the separation principle for the conventional optimal $\mathcal{H}_2$ output feedback design {which simplifies the output feedback control problem into a state feedback problem and a state observer design problem of the plant.}
However, the optimal state feedback gain here is based on the auxiliary plant \eqref{Equ:Plant_IID_II} derived from the models of the plant and the noise, and involves the input delay information of the plant, {rather than the original plant itself}.
This is conceptually different to the conventional separation principle.
On the other hand, the MARE \eqref{Equ:DARE_X_II} is a generalization of the MARE associated with optimal control of linear systems with i.i.d. multiplicative noise.
Both $\bar{A},\tilde{B}_2$ and $M(X)$ contain the information of the colored multiplicative noise.
}
\end{rem}

\section{Applications to NCSs}
\label{Sec:Applications}

As mentioned in Section \ref{Sec:Problem_Formulation}, the proposed multiplicative noise covers the classical i.i.d. multiplicative noise as well as the analog erasure channel.
In particular, this model can also be used to describe channel transmission delays such that the results are available to NCSs with random transmission delays.

\subsection{Channel with random transmission delays}
\label{Sec:NCS_Random_Delay}

\begin{figure}[hbt]
\centering
\subfloat[]{
\resizebox{0.49\linewidth}{!}{
\begin{tikzpicture}[auto, node distance=2cm, >=stealth', line width=0.7pt]
\small
\node[block](P){$P$};
\node[block, below right of= P, xshift=1cm, yshift=0.3cm](K){$K$};
\node[block, dashed, thin, below of= P, yshift=-0.425cm, minimum width=2.5cm, minimum height=1.7cm](channel){};
{\scriptsize
\node[sblock, below of= P, yshift=0.14cm](D0){$\mathbf{D}_{k,0}$};
\node[sblock, below of= P, yshift=-0.4cm](D1){$\mathbf{D}_{k,1}{\q}^{-1}$};
\node[sblock, below of= P, yshift=-0.97cm](D2){$\mathbf{D}_{k,2}{\q}^{-2}$}; %\delta(\tau_{k-2}\!-\!2)
}
\node[block, dashed, thin, below left of= P, xshift=-1.1cm, yshift=0.35cm, minimum height=1.3cm, minimum width=1.9cm](D){};
{\scriptsize
\node[sblock, below left of= P, xshift=-0.5cm, yshift=0.2cm, minimum width=0.5cm,](a0){$\alpha_0$};
\node[sblock, below left of= P, xshift=-1.1cm, yshift=0.2cm, minimum width=0.5cm,](a1){$\alpha_1$};
\node[sblock, below left of= P, xshift=-1.7cm, yshift=0.2cm, minimum width=0.5cm,](a2){$\alpha_2$};
\node[sum, above of= a1, yshift=-0.4cm](sum0){};
}
\draw[->]($(P.west)+(-2.5cm,0.2cm)$) -- node[near start]{$w$} ($(P.west)+(0cm,0.2cm)$);
\draw[->](sum0) |- node[]{$u_d$} ($(P.west)+(0cm,-0.2cm)$);
\draw[->]($(P.east)+(0cm,-0.2cm)$) -| node[near end]{$y$} (K);
\draw[->]($(P.east)+(0cm,0.2cm)$) -- node[near end]{$z$} ($(P.east)+(2.3cm,0.2cm)$);
\draw[->](K) |- node[near start]{$u$} (D1);
\draw[->]($(D1)+(1cm,0cm)$) |- (D0);
\draw[->]($(D1)+(1cm,0cm)$) |- (D2);
%\draw[->](channel) -| node[near end]{$\mathbf{v}_k$} (D);

\draw[->, thin](D0) -| ($(D1)+(-1cm,0.025cm)$) -|  ($(a1)+(0.025cm,-0.4cm)$) -| (a0); % node[near end, swap]{$\mathbf{v}$}
\draw[->, thin](D1) -| (a1);
\draw[->, thin](D2) -| ($(D1)+(-1cm,-0.025cm)$) -| ($(a1)+(-0.025cm,-0.4cm)$) -| (a2);

\draw[->, thin](a0) |- (sum0);
\draw[->, thin](a1) -- (sum0);
\draw[->, thin](a2) |- (sum0);
\path ($(channel) + (0,-1.1cm)$) node[rectangle](){\scriptsize 2-Step Delay Channel};
\path ($(D) + (-0.6cm,-0.75cm)$) node[rectangle](){\scriptsize Receiver};
\end{tikzpicture}
\label{Fig:System_model_example_a}
}
}
\subfloat[]{
\resizebox{0.45\linewidth}{!}{
\begin{tikzpicture}[auto, node distance=2cm, >=stealth', line width=0.7pt]
\small
\node[block](P){$P$};
\node[sum, left of= P, xshift=-0.5cm, yshift=-0.2cm](sum1){};
\node[block, below right of= P, xshift=0.5cm, yshift=0.3cm](K){$K$};
\node[block, below of= P](H){$H$};
\node[block, below of= H, yshift=0.5cm](D){$\varOmega$};

\draw[->]($(P.west)+(-2cm,0.2cm)$) -- node[pos=0.1]{$w$} ($(P.west)+(0cm,0.2cm)$);
\draw[->](sum1) -- node[]{} ($(P.west)+(0cm,-0.2cm)$);
\draw[->]($(P.east)+(0cm,-0.2cm)$) -|  node[near start]{$y$} (K);
\draw[->]($(P.east)+(0cm,0.2cm)$) -- node[pos=0.99]{$z$} ($(P.east)+(2.5cm,0.2cm)$);
\draw[->](K) |- (H);
\draw[->](H) -| node[near end, swap]{$\bar{u}$} (sum1);
\draw[->](K) |-  node[pos=0.2]{$u$} (D);
\draw[->](D) -- ++(-2.5cm,0cm) |- node[pos=0.2]{$d$} (sum1);

\path (sum1.west |- P.north)+(-0.3,0.2) node (con1) {};
\path (H.south -| K.east)+(0.2,-0.2) node (con2) {};
\path [dashed, block] (con1) rectangle (con2);

\node[right of= P, xshift=0.5cm, yshift=0.5cm]{$G$};
\end{tikzpicture}
\label{Fig:System_model_example_b}
}
}
\caption{Feedback system over a 2-step random delay channel}
\label{Fig:System_model_example}
\end{figure}
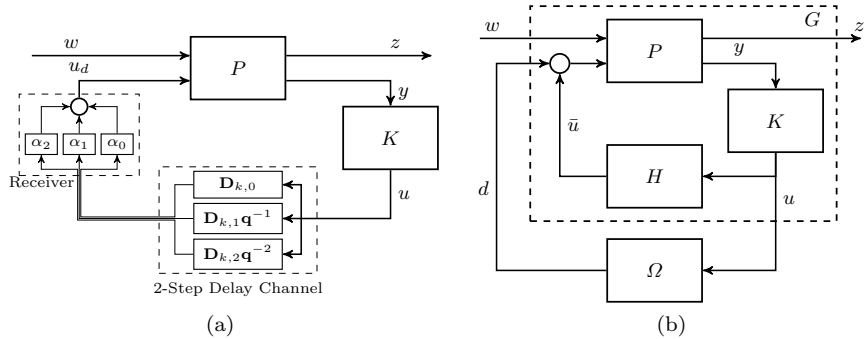

For simplicity, a 2-step random delay channel is taken into consideration.
As depicted in Fig. \ref{Fig:System_model_example_a}, the communication channel with random delay is placed in the path from the LTI controller $K$ to the receiver, where ${\q}^{-1}$ is the backward shift operator and
\begin{align*}
  \mathbf{D}_{k,i} &= \delta(\tau_{k-i}-i),~~i=0,1,2,
\end{align*}
with $\tau_k$ being the channel induced delay for the control signal $u(k)$ and taking values from the set $\bar{\mathcal{D}}=\{0, 1, 2\}$.
Note that the function $\mathbf{D}_{k,i}$ can indicate that whether the signal $u({k-i})$ arrives the receiver at time $k$ and guarantee that the control signal would arrive the receiver only once.
Here, data disordering is allowed in the network such that a {linear combination} of the received data at time $k$ is taken as the output of the receiver, i.e.,
\begin{align}\label{Equ:w_k}
u_d(k) = \sum_{i=0}^{2}\alpha_i \delta(\tau_{k-i}-i)u({k-i})
\end{align}
where the weights $\alpha_0,\alpha_1,\alpha_{2}$ are assigned to the received signals according to their delay steps, respectively, provided that the transmitted data are time-stamped.

It would be natural to assume that $\{\tau_k: k\in \mathbb{Z}\}$ is an i.i.d. random process with a probability mass function
\begin{align*}%\label{Equ:Prob_Mass_Func}
p_i=\Pr \left\{ {{\tau_k} = i} \right\},~~i=0,1,2
\end{align*}
where $p_i \in [0,1]$ and \hlt{$p_0 + p_1 + p_2 = 1$}.
From the stochastic properties of $\tau_k$, for any $k$ it holds that $\E\{\delta(\tau_{k-i}-i)\} =  p_i$.
Under this condition, it can be shown that the transmission delay induced multiplicative noise described by \eqref{Equ:w_k} satisfies Assumption \ref{Assum:omega_k} with
\begin{align*}
\omega(k,k-i) = \alpha_i \delta(\tau_{k-i}-i)
\end{align*}
so that
$\mu_i = \alpha_i p_i$ and $\beta_{i,j}= \delta(i-j) \alpha_i^2 p_i - \alpha_i \alpha_j p_i p_j$.
This means that the results obtained in the previous section can be applied to the NCS shown in Fig. \ref{Fig:System_model_example} directly.
More precisely, $u_d(k)$ can be written as $u_d(k) = H({\q}^{-1})u({k}) + \varOmega({\q}^{-1}) u({k})$,
where $H({\q}^{-1}) = \sum_{i=0}^{2}\alpha_i p_i {\q}^{-1}$ is the mean system and
\begin{align*} %\label{Equ:Delta_examp}
\varOmega({\q}^{-1}) = \sum_{i=0}^{2}\alpha_i [\delta(\tau_{k-i}-i) - p_i]{\q}^{-i}
\end{align*}
is the delay-induced uncertainty.
Consequently,  Fig. \ref{Fig:System_model_example_a} can be rewritten as  Fig. \ref{Fig:System_model_example_b}, where $G$ is the nominal system in \eqref{Equ:Ge}.
It shows that the energy spectral density of the channel-induced uncertainty $\varOmega$ is given by \cite{Su2017mean-square}
\begin{align*}
S(z)=\frac{1}{2} \sum_{i,j\in \bar{\mathcal{D}}}(\alpha_{i}z^{i}-\alpha_{j}z^{j})
(\alpha_{i}z^{-i}-\alpha_{j}z^{-j})p_{i}p_{j} % \label{Equ:S_special}
\end{align*}
which implies that the spectral factor $\varPhi(z)$ exists.
Note that this case can be generalized to $\bar{\tau}$-step random delay channel with $\bar{\tau}$ being any integer.

\subsection{Analog erasure channel}

%To recover the result in analog erasure channel,
Consider the state feedback design for \hlt{minimizing average control power problem} of a plant $P$ which admits a stabilizable state-space realization that
\begin{align*}
x(k + 1) &= A x(k) + B (w(k) + u_d(k)),~~~x(0)=0 \\
z(k) &= \hlt{ u_d (k)}
\end{align*}
Suppose now, in Fig. \ref{Fig:System_model_example_a}, an analog erasure channel is placed between the controller output $u$ and the system input $u_d$ that
%\begin{align*}
$u_d(k) = \theta_k u(k)$,
%\end{align*}
where $\{\theta_k: k \in \mathbb{Z}\}$ is an i.i.d. Bernoulli process satisfying $\Pr\{\theta_k=0\}=e$ and $\Pr\{\theta_k=1\}=1-e$.
Define $\varGamma_{e} = \frac{1-e}{e}$. %be the generalized signal-to-noise ratio (SNR) constraint.

\begin{cor}\label{Cor:Optimal_cost_erasure}
Suppose $\lambda_i,i=1,\cdots,n$ are the unstable poles of $P$.
The closed-loop networked feedback system is mean-square stable if and only if
\begin{align}
\frac{1}{2}\log\left(1+\varGamma_{e}\right) > \sum_{i=1}^{n}\log|\lambda_i| \label{Equ:Gamma_SNR1}
\end{align}
or, equivalently, \hlt{$e\!<\!\mathcal{M}^{-2}$, where $\mathcal{M} = \prod_{i=1}^{n}|\lambda_i|$.
If the mean-square stability criterion holds, the minimum average control power of the system is given by
\begin{align}\label{Equ:Optimal_J1}
%\hlt{\min_{K \in \mathcal{K}}\pow{u_d}^2 =  \frac{(\varGamma_e+1)\left(\prod_{i=1}^{n}|\lambda_i|^2 - 1\right)}{\varGamma_{e} - \left( \prod_{i=1}^{n}|\lambda_i|^2 - 1 \right)}}.
\min_{K \in \mathcal{K}}\pow{u_d}^2 =  \frac{\mathcal{M}^2 - 1}{1- e \mathcal{M}^2}.
\end{align}} %is the Mahler measure of the system matrix $A$.
\end{cor}

\begin{pf}
Note that the analog erasure channel is a special case of the proposed noise without memory.
That is, $\bar{\tau} = 0$ with $\mu_0 = 1-e$ and $\beta_0 = e(1-e)$.
Then the mean system and spectral factor of the uncertainty are $H = 1-e$ and $\varPhi = \sqrt{e(1-e)}$, respectively, such that $\bar{A} = A,~\bar{B}_1 = B,~\bar{B}_2 = \sqrt{e(1-e)} B,~\tilde{B}_2 = (1-e) B$.
\hlt{Then the MARE \eqref{Equ:DARE_X_II} becomes
\begin{align}
X = A^T X A \!-\! \mu_0 { A^T X {{B}} ({{  1 +  {{B}^T}X{{B}}}})^{-1} {{B}^T}X A },  \label{Equ:MDARE}
\end{align}
%This lead to a classical MARE in networked control area,
whose associated optimal performance is given by $B^TX B$.
%\begin{align*}
%J_{opt} = B^TXB.
%\end{align*}
Since $B$ is rank one, it is well-known that the above equation has a nonzero positive semi-definite solution if and only if $\mu_0 > 1 - \mathcal{M}^{-2}$ \cite{Schenato2007Foundations},
%\begin{align*}
%\mu_0 > 1 - \frac{1}{\prod_{i=1}^{n}|\lambda_i|^2},
%\end{align*}
which, by simple calculation, is equivalent to \eqref{Equ:Gamma_SNR1}.
To prove \eqref{Equ:Optimal_J1}, we rewrite the classical MARE above as
\begin{align}
X = A^T X A -  { A^T X {{B}} ({{ \beta +  {{B}^T}X{{B}}}})^{-1} {{B}^T}X A }, \label{Equ:MDARE_2}
\end{align}
where $\beta = \mu_0^{-1} + (\mu_0^{-1}-1){{B}^T}X{{B}}$.
Let $X_{\beta} = X\beta^{-1}$, then the equation \eqref{Equ:MDARE_2} becomes a standard DARE for optimal average control power problem of a system, whose optimal cost is given by $B^T X_{\beta} B = { \prod_{i=1}^{n}|\lambda_i|^2 - 1}$, see \cite{Elia2005Remote,Elia2001Stabilization}.
%\begin{eqnarray*}
%B^T X_{\beta} B = { \prod_{i=1}^{n}|\lambda_i|^2 - 1}.
%\end{eqnarray*}
Therefore, the optimal cost for the original plant is
%\begin{align*}
%B^T X B &=
%({\mu_0^{-1} + \varGamma_e^{-1} {{B}^T}X{{B}}} ) \Big( { \prod_{i=1}^{n}|\lambda_i|^2 - 1} \Big)
%\end{align*}
$B^T X B =
({\mu_0^{-1} + \varGamma_e^{-1} {{B}^T}X{{B}}} ) ( { \prod_{i=1}^{n}|\lambda_i|^2 - 1} )$
since $\varGamma_e = ({1-e})/{e}$.}
Solving for $B^T X B$ gives \eqref{Equ:Optimal_J1} and completes the proof.
%\begin{align*}
%B^T X B &=
%\frac{ \prod_{i=1}^{n}|\lambda_i|^2 - 1}{(1-e)^2 - e(1-e)\left( \prod_{i=1}^{n}|\lambda_i|^2 - 1 \right)},
%\end{align*}
%which completes the proof.
\end{pf}

\hlt{
\begin{rem}
Corollary \ref{Cor:Optimal_cost_erasure} recovers the famous results of maximum erasure rate and linear-quadratic/$\mathcal{H}_2$ optimal control on systems with analog erasure channel, see \cite{Schenato2007Foundations,Elia2011}.
It shows that the MARE \eqref{Equ:DARE_X_II} is reduced to \eqref{Equ:MDARE} which is well-known in studying estimation and optimal control problems of systems with analog erasure channel \cite{SINOPOLI2005LQG}.
Meanwhile, the equality \eqref{Equ:Optimal_J1} provides an explicit form of the minimum average (corrupted) control power, which reveals the effect of the erasure rate on the average control power: when $e$ approaches 0, \eqref{Equ:Optimal_J1} reduces to the classical minimum average control power \cite{Elia2011}, and the optimal average control power increases with the erasure rate.
\end{rem}
}

\section{Numerical example}
\label{Sec:Simulation}

\begin{figure}[!htb]
  \centering
\subfloat[Convergent optimal performance]{
  \includegraphics[width=7.5cm]{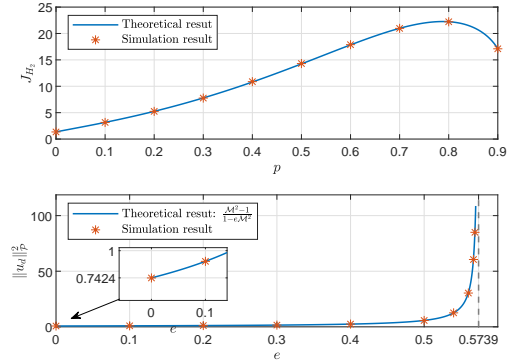}
  %\caption{Optimal control performances}
  \label{Fig:Ezz}
}\\ \vspace{-0.3cm}
\subfloat[Plant state and estimate in one simulation ($\epsilon=0.8,~p=0.3$)]{
\includegraphics[width=7.5cm]{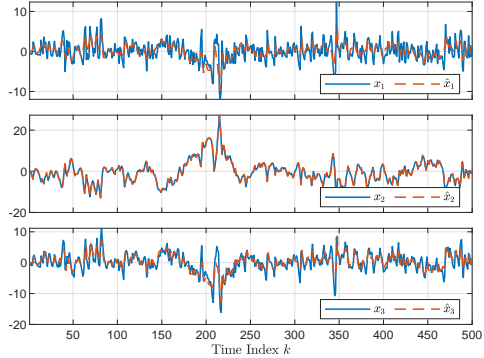}
  %\caption{Optimal control performances}
  \label{Fig:x_est}
}
\caption{Optimal performances}\label{Fig:optimal_performance}
\end{figure}

\hlt{
In this section, a networked system with random transmission delay shown in Section \ref{Sec:Applications} is proposed to verify the results obtained.
Consider a discrete-time LTI plant $P_\epsilon$ with state-space realization:
\begin{align}
\hspace{-0.7cm}
x(k + 1) &= \begin{bmatrix}\begin{smallmatrix}
               1.1 & 0 & 0 \\
               1 & 1.2 & 0 \\
               1 & 0   & 0.5
             \end{smallmatrix}\end{bmatrix} x(k) + \begin{bmatrix}\begin{smallmatrix}
                                    1 \\
                                    0.5\epsilon \\
                                    1
                                  \end{smallmatrix} \end{bmatrix} w(k) + \begin{bmatrix}\begin{smallmatrix}
                                                         1 \\
                                                         0 \\
                                                         1
                                                       \end{smallmatrix} \end{bmatrix} u_d(k)\nonumber\\
z(k) &= \begin{bmatrix}\begin{smallmatrix}
           0 & \epsilon & 2\epsilon
         \end{smallmatrix}\end{bmatrix} x(k) + u_d(k)  \label{Exam:P}\\
y(k) &= \begin{bmatrix}\begin{smallmatrix}
           1 & 0 & 1\\
           0 & 1 & 0
        \end{smallmatrix} \end{bmatrix} x(k) \nonumber
\end{align}where $\epsilon$ is a real number.
It can be shown that the relative degrees of the transfer functions from $w$ and $u_d$ to $y$ are $r_1 = 1$ and $r_2 = 1$, respectively.
Note that $\mathcal{M}={\prod_{\lambda_i \in \mathbb{D}^c}|\lambda_i|} = 1.32$.
Here consider that the colored multiplicative noise is induced by the 2-step random delay channel with delay probabilities
\begin{align*}
p_0 = 0.9-p,~~p_1 = p,~~p_2 = 0.1.
\end{align*}
where $0\le p \le 0.9$.
%We restrict $\epsilon$ to $[0,1]$ and
Suppose the weights of the received data are given \emph{a priori} with $\alpha_0 = 1,~\alpha_1  = 0.67$, and $\alpha_2 = 0$.
%The case $\alpha_2 = 0$ represents that the communication channel is with packet loss rate $0.1$.
For $P_\epsilon$ with $\epsilon \ne 0$, the problem of minimizing \eqref{Equ:J_H2} is directly solved by Theorem \ref{Thm:K_opt_L_L0}. %when $\epsilon=0$, it is reduced to an \hlt{optimal average control power} problem for SISO networked system with analog erasure channel.
%If the latter is the case, under the above setting, it holds that
%\begin{align*}
%0.4 = e = p_1+p_2 < {\prod_{\lambda_i \in \mathbb{D}^c}|\lambda_i|^{-2}} = 0.5739
%\end{align*}
%so that the closed-loop system is mean-square stabilizable and the optimal control cost is given by
%\hlt{\begin{align*}
%\hspace{-0.5cm}\min_{K \in \mathcal{K}}\pow{u_d}^2 = \frac{ (\varGamma_e+1) \big(\prod_{\lambda_i \in \mathbb{D}^c}|\lambda_i|^2 - 1\big)}{\varGamma_{e} - \big( \prod_{\lambda_i \in \mathbb{D}^c}|\lambda_i|^2 - 1 \big)}= 2.4498.
%\end{align*}}
Take $\epsilon = 0.8$ as an example, it shows that the positive semi-definite solution to the MARE \eqref{Equ:DARE_X_II} exists (which is omitted due to space limit) for all $p\in [0,0.9]$ such that the networked system is mean-square stabilizable. %and the optimal control performance is given by $J_{opt}=7.7941$.
Fig. \ref{Fig:Ezz} shows the theoretical and simulated results of the optimal performance for the plant $P_{\epsilon = 0.8}$ with random transmission delay channel, over 20000 Monte-Carlo runs.
It shows that the simulated result perfectly coincides with the theoretical one, with the percent errors less than $0.2\%$.
With the increase of the probability of one-step delay, the optimal performance increases except when the probability approaches one, which implies the non-convex relation between the delay probability and the optimal performance.
Since the controller is observer-based, the state of the plant and the associated estimate in one out of the simulations are illustrated in Fig. \ref{Fig:x_est} for $p=0.3$.

By resetting $\alpha_1 = \alpha_2 = 0$ (i.e., the delayed data are actively dropped in the receiver), the networked system reduces to the case of analog erasure channel with erasure rate $e = p_1 + p_2$.
Here, $0\le e \le 1$ is taken into consideration.
Corollary \ref{Cor:Optimal_cost_erasure} shows that if $e < 0.5739 = \mathcal{M}^{-2}$ holds,
%\begin{align*}
%e < 0.5739 = \mathcal{M}^{-2}
%\end{align*}
then the closed-loop system is mean-square stabilizable and the minimum average control power for the plant $P_{\epsilon=0}$ via state feedback is explicitly given by \eqref{Equ:Optimal_J1}.
%\begin{align*}
%\min_{K \in \mathcal{K}}\pow{u_d}^2 =  \frac{\mathcal{M}^2 - 1}{1- e \mathcal{M}^2} = 2.4498.
%\end{align*}
The theoretical and simulated results of the optimal average control power are depicted in Fig. \ref{Fig:Ezz}.
Notice that the minimum average control power is $\mathcal{M}^2 - 1 = 0.7424$ if the erasure channel is absent, and when the erasure rate $e$ approaches $\mathcal{M}^{-2}$, large control effort is required to stabilize the system.
}

\section{Conclusion}
\label{Sec:Conclusion}

In this paper, the mean-square optimal control problem of a linear feedback system with a quasi-colored multiplicative noise via output feedback is addressed.
It shows that the noise under study generalizes the classical i.i.d. multiplicative noise and has advantage on modeling a class of network phenomena including random transmission delays.
By converting the optimal control problem into a mean-square stability problem of an augmented system, the optimal output feedback controller is designed in terms of a mean-square stabilizing solution to one MARE.
The necessary and sufficient condition for the existence of the mean-square stabilizing solution to the MARE is also presented.
Owing to the feature of the proposed multiplicative noise, the result is applied to optimal control problems of NCSs with random transmission delays and analog erasure channel, respectively.
The methodology would be expected to provide a novel idea to deal with a class of channel-induced uncertainties in NCSs.

\hlt{
It is worthwhile to figure out that one can refer to \cite{Su2017mean-square} for solving the multiple external input case associated with this work.
%Note that the external input $w$ and the control input $u_d$ of the plant \eqref{Equ:system} is scalar-valued in this work.
Inspired by \cite{Xiao2012Feedback,Li2018Stability}, a multi-dimensional counterpart of the proposed colored-noise would be straightforward.
Thus, the multivariable case of the optimal control problem in this work would be expected to be solved by applying the proposed method and focus of future research.
}

\appendix
\section{Proof of Lemma \ref{Lma:J_H2}}\label{Appendix:L7}

Note that for the system in Fig. \ref{Fig:G_with_Omega} we have
\begin{align*}
z(k) &= G_{zw}(\q)w(k) + G_{zd}(\q)d(k) \\
u(k) &= G_{uw}(\q)w(k) + G_{ud}(\q)d(k)
\end{align*}
where ${\q}^{-1}$ is the shift operator.
Since $w(k)$ is assumed to be zero-mean and unit-variance and independent of $d(k)$,
\begin{align}
\pow{z}^2 &= \|G_{zw}\|_2^2 + \pow{G_{zd}(\q)d(k)}^2 \label{Equ:z_pow} \\
\pow{u}^2 &= \|G_{uw}\|_2^2 + \pow{G_{ud}(\q)d(k)}^2 \label{Equ:u_pow}
\end{align}
It is claimed that for any stable and proper LTI \hlt{single-input} plant $G_0$, it holds $\pow{G_0(\q)d(k)}^2 = \|G_0(z)\varPhi(z)\|_2^2 \pow{u}^2$.
Let $\{g_0(l)\}_{l=0}^\infty$ be the unit-impulse response of $G_0$ such that the output of $G_0$ to the input $d(k)$ is given by $z_0(k) = \sum_{i=0}^{k}g_0(i)d(k-i)$.
Suppose $d(k)=0$ for $k<0$, it shows that
\begin{align*}
\hspace{-0.6cm}\E\{\|z_0(k)\|^2\} % &= \E\left\{ \sum_{l=0}^{\infty}\sum_{m=0}^{\infty}g_0(l)d(k\!-\!l)g_0(m)d(k\!-\!m) \right\}\\
 &=  \sum_{l=0}^{\infty}\sum_{m=0}^{\infty}g_0^T(l)g_0(m)\E\{d(k-l)d(k-m)\}.
\end{align*}
From \eqref{Equ:d_total} and Assumption \ref{Assum:omega_k}, it holds that
\begin{align*}
&\E\{d(k-l)d(k-m)\}= \sum_{i=0}^{\bar{\tau}} \beta_{i,l-m+i} \E\{u^2(k-l-i)\} %\\
%&= \sum_{i=0}^{\bar{\tau}} \sum_{j=0}^{\bar{\tau}}\delta ((k\!-\!l\!-\! i)-(k\! -\! m \!-\!j)) \beta_{i,j}  \E\{u^2(k\!-\!l\!-\!i)\}\\
%&= \sum_{i=0}^{\bar{\tau}} \beta_{i,l-m+i} \E\{u^2(k-l-i)\}.
\end{align*}
Thus,
\begin{align*}
\pow{z_0}^2 &=  \lim_{\bar{k} \to \infty } \frac{1}{{\bar{k} + 1}}\sum\limits_{k = 0}^{\bar{k}} \sum_{l=0}^{\infty}\sum_{m=0}^{\infty} g_0^T(l)g_0(m) \\
&\hspace{1.5cm} \times\sum_{i=0}^{\bar{\tau}} \beta_{i,l-m+i} \E\{u^2(k-l-i)\}\\
&= \left( \sum_{l=0}^{\infty}\sum_{m=0}^{\infty} g_0^T(l)g_0(m)\sum_{i=0}^{\bar{\tau}} \beta_{i,l-m+i} \right)\pow{u}^2\\
&=  \left( \sum_{l=0}^{\infty}\sum_{j=-\infty}^{\infty} g_0^T(l)g_0(l-j) r(j) \right) \pow{u}^2.
%&= \|G_0(z)\varPhi(z)\|_2^2 \pow{u}^2
\end{align*}
By the definition of $\mathcal{H}_2$ norm, it shows that
\begin{align*}
&\sum_{l=0}^{\infty}\sum_{j=-\infty}^{\infty} g_0^T(l)g_0(l-j) r(j)= \|G_0(z)\varPhi(z)\|_2^2,
\end{align*}
provided that $\hlt{\varPhi(z)\varPhi(z^{-1}) = S(z)}$. %, the energy spectral density of the uncertainty $\varOmega$.
As a result,
\begin{align*}
\pow{G_{zd}(\q)d(k)}^2 &= \|G_{zd}(z)\varPhi(z)\|_2^2 \pow{u}^2\\
\pow{G_{ud}(\q)d(k)}^2 &= \|G_{ud}(z)\varPhi(z)\|_2^2 \pow{u}^2
\end{align*}
Consequently, followed by \eqref{Equ:z_pow} and \eqref{Equ:u_pow}, $\pow{u}^2$ and $\pow{d}^2$, as well as $\E\{u^2(k)\}$ and $\E\{d^2(k)\}$, exist if and only if $\|G_{ud}(z)\varPhi(z)\|_2^2 < 1$
%\begin{align*}
%\|G_{ud}(z)\varPhi(z)\|_2^2 < 1,
%\end{align*}
and, if this is the case, Lemma \ref{Lma:J_H2}.2 holds by solving $\pow{z}^2$. %, then
%\begin{align*}
%\hspace{-0.2cm}J_{H_2} = {\left\| {{G_{zw}}} \right\|_2^2 + {{\left\| {{G_{uw}}} \right\|_2^2 ({1 - \| G_{ud} \varPhi\|_2^2})^{-1} \| G_{zd} \varPhi \|_2^2}}}.
%\end{align*}
%These complete the proof.
Lemma \ref{Lma:J_H2}.3 follows from the same technique in \cite{Lu2002MeanSquare} together with Lemma \ref{Lma:J_H2}.2.
One also can directly compute the spectral radius of $\hat{G}$ to obtain the result since $\hat{G}$ is a $2\times 2$ matrix.

\section{Proof of Lemma \ref{Lma:P_gamma_order_reduced_realization}}\label{Appendix:P_gamma}

It is standard to rewrite $P_\gamma$ into a form of lower linear fractional transformation that $P_\gamma = \mathcal{F}(\bar{P}_\gamma,K)$ for some plant $\bar{P}_\gamma$.
Then a direct realization of the auxiliary plant $\bar{P}_\gamma(z)$ is given by
%\begin{small}
\begin{align}
\varGamma^{\frac{1}{2}} \left[\begin{array}{ccc|ccc}
                      A & B_2 \hat{C} & B_2 \hat{C} & B_1 & B_2 \hat{D}_2 & B_2 \hat{D}_1 \\
                      0 & \hat{A} & 0 & 0 & 0 & \hat{B}_1 \\
                      0 & 0 & \hat{A} & 0 & \hat{B}_2 & 0 \\ \hline
                     C_1 & D \hat{C} & D \hat{C} & 0 & D \hat{D}_2 & D \hat{D}_1 \\
                     0 & 0 & 0 & 0 & 0 & 1 \\
                     C_2 & 0 & 0 & 0 & 0 & 0
                    \end{array}\right]\varPi \label{Equ:P_Gamma}
\end{align}
%\end{small}
By applying the algebraic equivalence transformation $T = \left[\begin{smallmatrix}
                                                         I & 0 & 0 \\
                                                         0 & I & -I \\
                                                         0 & 0 & I
                                                       \end{smallmatrix}\right]$,
%\begin{align*}
%T = \begin{bmatrix}
%                                                         I & 0 & 0 \\
%                                                         0 & I & -I \\
%                                                         0 & 0 & I
%                                                       \end{bmatrix}
%\end{align*}
%to the above realization,
we can see that \eqref{Equ:P_Gamma} can be reduced to \eqref{Equ:Auxiliary_Plant_1}, without affecting the stabilizability and detectability of the plant since the reduced mode is related to $\hat{A}$.

\section{Proof of Lemma \ref{Lma:Optimal_for_auxiliary_plant}}\label{Appendix:AB_AbarBbar}

To see the existence of solutions to the DAREs \eqref{Equ:Riccati_Equ_X} and \eqref{Equ:Riccati_Equ_Y}, it is sufficient to show that i) $(\bar{A},\tilde{B}_{2})$ is stabilizable and $(\bar{A},\bar{C}_{1})$ has no unobservable poles on the unit circle, and ii) $(\bar{A},\bar{C}_{2})$ is detectable and $(\bar{A},[\bar{B}_{1}~\bar{B}_{2}])$ has no unstabilizable poles on the unit circle.
Suppose that Assumption \ref{Assump:A_B_stabilizable} holds.
It is easy to see that $(A,C_1)$ and $(\bar{A},\bar{C}_1)$ share the same unobservable poles.
Now we need to show that $(\bar{A},\tilde{B}_{2})$ is also stabilizable.
Otherwise, there exists an nonzero vector $\bar{v}^* = \begin{bmatrix}
                                                 v^* & \hat{v}^*
                                               \end{bmatrix}$ and $z_0$ with $|z_0|>1$ such that $\bar{v}^*\begin{bmatrix}
     z_0I-\bar{A} & \tilde{B}_{2}
   \end{bmatrix} =0$.
%\begin{align*}
%&\bar{v}^*\begin{bmatrix}
%     z_0I-\bar{A} & \bar{B}_2
%   \end{bmatrix} =0
%%&=\begin{bmatrix}
%%                     v^*(z_0I\!-\!A) & {v}^*B_2\hat{C} \!+\! \hat{v}^*(z_0I\!-\!\hat{A}) & v^*B_2 \hat{D}_2 \!+\! \hat{v}^*\hat{B}_1
%%                   \end{bmatrix} \\
%%&= 0.
%\end{align*}
By algebraic computation, it holds that $v^*(z_0I-A) = 0$ and $v^* B_2 \hat{C}(z_0I-\hat{A})^{-1}\hat{B}_1 = 0$.
%\begin{align*}
%v^*(z_0I-A) = 0 \text{ and } v^* B_2 \hat{C}(z_0I-\hat{A})^{-1}\hat{B}_1 = 0.
%\end{align*}
Notice that $\hat{C}(z_0I-\hat{A})^{-1}\hat{B}_1 = H(z_0) \ne 0$ as $z_0$ is an unstable pole of $P$, then $v^*B_2 = 0$, which implies that $(A,B_2)$ is unstabilizable and leads to a contradiction.
The converse is similar.
Note that ii) also can be obtained by similar technique, with the fact that $\varPhi(z)$ is minimum phase, and thus is omitted here.
The rest is standard for optimal control of discrete-time LTI systems, see \cite{Chen1995Optimal}.

\section{Proof of Lemma \ref{Lma:L_L0_J}}\label{Appendix:L_L0}

%Firstly, we need to show, for the auxiliary plant \eqref{Equ:Auxiliary_Plant_1}, that the relative degrees from $w$ and $\varphi$ to $y$ are respectively identical to $r_1$ and $r_2$.

Directly computing yields that the relative degree of $(\bar{A},\bar{B}_{1},\bar{C}_{2})$ is $r_1$.
For $i=0,1,\cdots$,
%\begin{align*}
$\bar{C}_{2}\bar{A}^i \bar{B}_{2} \!=\! (C_2 A^i B_2) \hat{D}_2 \!+\! \sum_{j=1}^{i} (C_2 A^{i-j} B_2) \hat{C} \hat{A}^{j-1} \hat{B}_{2}$.
%\end{align*}
Since $\hat{D}_2 \ne 0$, that the relative degree of the system $(A,B_2,C_2)$ is $r_2$ yields $\bar{C}_{2}\bar{A}^i \bar{B}_{2} = 0$ for $i=0,\cdots,r_2-2$ and $\bar{C}_{2}\bar{A}^{r_2-1} \bar{B}_{2} \ne 0$.
That is, the relative degree of $(\bar{A},\bar{B}_{2},\bar{C}_{2})$ is $r_2$.
It can be shown that $\bar{C}_2 \bar{\varPsi} = C_2 \varPsi$ so that $\bar{C}_2 \bar{\varPsi}$ is full column rank.
Meanwhile, it follows from \cite{Silverman1976Discrete} that, by Assumption \ref{Assum:Relative_degree}, for any given $\lambda_0,\lambda_1$, and $\gamma$, the stabilizing solution to the DARE \eqref{Equ:Riccati_Equ_Y} is given by
\begin{align*} %\label{Equ:Y}
Y_{\gamma} \!=\! \sum_{i=0}^{r_1 - 1}\bar{A}^i \bar{B}_{1}\sigma_0^2 \lambda_0^2 \bar{B}_{1}^T \bar{A}^{T^i} \!+\! \sum_{i=0}^{r_2 - 1}\bar{A}^i \bar{B}_{2}  {\lambda_1^2}{\gamma^{-2}} \bar{B}_{2}^T \bar{A}^{T^i}.
\end{align*}
Thus, $Y_\gamma \bar{C}_2^T = \bar{\varPsi}\varPi^2 \bar{\varPsi}^T \bar{C}_2^T$, which yields
%\begin{align*}
$Y_\gamma \bar{C}_2^T (\bar{C}_2 Y_\gamma \bar{C}_2^T)^\dagger = \bar{\varPsi} (\bar{C}_2 \bar{\varPsi})^\dagger$.
%\end{align*}
%$Y_\gamma \bar{C}_2^T (\bar{C}_2 Y_\gamma \bar{C}_2^T)^\dagger = \bar{\varPsi} (\bar{C}_2 \bar{\varPsi})^\dagger$.
Then \eqref{Equ:Optimal_L1} holds.
%Note that
%\begin{align*}
%\hspace{-0.6cm}&(L_{0\gamma}\bar{C}_{2} - F_{\gamma})Y_{\gamma}(L_{0\gamma}\bar{C}_{2} - F_{\gamma})^T\\
%&= F_\gamma [Y_\gamma - Y_\gamma \bar{C}_2^T (\bar{C}_2 Y_\gamma \bar{C}_2^T)^\dagger \bar{C}_2 Y_\gamma ] F_\gamma^T\\
%&= F_\gamma (Y_\gamma - \bar{\varPsi} \varPi^2 \bar{\varPsi}^T ) F_\gamma^T.
%\end{align*}
It is observed that $(L_{0\gamma}\bar{C}_{2} - F_{\gamma})Y_{\gamma}(L_{0\gamma}\bar{C}_{2} - F_{\gamma})^T = F_\gamma (Y_\gamma - \bar{\varPsi} \varPi^2 \bar{\varPsi}^T ) F_\gamma^T$.
Then
\begin{align*}
&\hspace{-0.8cm}F_\gamma (Y_\gamma - \bar{\varPsi} \varPi^2 \bar{\varPsi}^T ) F_\gamma^T = R_\gamma \bigg(\sum_{i=0}^{r_1 - 2}F_\gamma\bar{A}^i \bar{B}_{1}\sigma_0^2 \lambda_0^2 \bar{B}_{1}^T \bar{A}^{T^i}F_\gamma^T \\
&\hspace{1.5cm}+ \sum_{i=0}^{r_2 - 2}F_\gamma\bar{A}^i \bar{B}_{2}  {\lambda_1^2}{\gamma^{-2}} \bar{B}_{2}^T \bar{A}^{T^i}F_\gamma^T\bigg) R_\gamma^T\\
&=\sigma_0^2 \lambda_0^2 \sum_{i=0}^{r_1 - 2}\bar{B}_{1}^T \bar{A}^{T^i} (F_\gamma^T R_\gamma^TR_\gamma F_\gamma)\bar{A}^i \bar{B}_{1} \\
&\hspace{1.5cm}+ {\lambda_1^2}{\gamma^{-2}} \sum_{i=0}^{r_2 - 2}\bar{B}_{2}^T \bar{A}^{T^i}(F_\gamma^T R_\gamma^TR_\gamma F_\gamma)\bar{A}^i \bar{B}_{2}.
\end{align*}
Note that  $F_\gamma^T R_\gamma^T R_\gamma F_\gamma = \bar{A}^T X_{\gamma} \bar{A} + \bar{C}_{1}^T \bar{C}_{1} - X_{\gamma}$.
%\begin{align*}
%&F_\gamma^T R_\gamma^T R_\gamma F_\gamma = \bar{A}^T X_{\gamma} \bar{A} + \bar{C}_{1}^T \bar{C}_{1} - X_{\gamma}.
%\end{align*}
Then \eqref{Equ:Optimal_J_1} holds.

\section{Proof of Lemma \ref{Lma:Optimal_F_J}}\label{Appendix:Jh2_F}

The former part and the optimal state feedback gain can be obtained by applying \cite[Lemma 12]{Su2016Control}.
Then the associated optimal cost is given by
\begin{align*}
&\min_{F}\tilde{J}_{H_2} = \phi_0(X) \\
&= {\bar{B}_{1}^T}\bar{A}^{T^{r_1-1}}{X} \bar{A}^{{r_1-1}} {\bar{B}_{1}} +\sum\nolimits_{j=0}^{r_1- 2}{\bar{B}_{1}^T}\bar{A}^{T^{j}}\bar{C}_{1}^T \bar{C}_{1} \bar{A}^{j} {\bar{B}_{1}}
\end{align*}
Since $\bar{A}^i\bar{B}_1 = \begin{bmatrix}\begin{smallmatrix}
                              A^i B_1 \\
                              0
                            \end{smallmatrix}\end{bmatrix}$ for any integer $i$,
the optimal cost in \eqref{Equ:Optima_Jh2_1} is obtained.

%\printcredits

%\bibliographystyle{IEEEtran}
%\bibliography{optimal-ref}

\end{document}